\documentclass{article}

\usepackage{arxiv}
\usepackage[utf8]{inputenc}
\usepackage[english]{babel}
\usepackage[title]{appendix}

\usepackage{subfig}
\usepackage{newtxtext,newtxmath}
\usepackage{wrapfig}
\usepackage{pdfpages}
\usepackage{titlesec}
\usepackage[tikz]{mdframed}
\usepackage{diagbox}
\usepackage{relsize}
\usepackage{xurl}

\usepackage{geometry} 

\usepackage[utf8]{inputenc}
\usepackage{mathtools}
\usepackage{tabularx}
\newcolumntype{C}[1]{>{\centering\arraybackslash}p{#1}}
\usepackage{multirow}
\usepackage{tabularx}
\usepackage{makecell}
\usepackage{siunitx}
\usepackage{adjustbox}
\usepackage{chngcntr}
\usepackage{caption} 
\usepackage{comment}
\usepackage{tcolorbox}
\usepackage{bigdelim}

\usepackage{caption}
\usepackage{algorithm}
\usepackage{algpseudocode}

\usepackage{booktabs}

\newcolumntype{Y}{>{\raggedleft\arraybackslash}X}

\DeclareUnicodeCharacter{2212}{-}

\usepackage{setspace}

\makeatletter
\newcommand{\mylabel}[2]{%
   \protected@write \@auxout {}{\string \newlabel {#1}{{#2}{\thepage}{#2}{#1}{}} }%
   \hypertarget{}{}
}
\makeatother



\usepackage{wrapfig}

\usepackage{amsmath}

\DeclareMathOperator*{\argmin}{arg\,min}

\newif\ifdraft
\draftfalse
\drafttrue
\ifdraft
  \newcommand{\ian}[1]{{\textcolor{red}{ Ian: #1 }}}
  \newcommand{\liz}[1]{{\textcolor{magenta}{ Liz: #1 }}}
  \newcommand{\tak}[1]{{\textcolor{black}{#1}}}
 \newcommand{\takbefore}[1]{{\textcolor{magenta}{ TakBefore: #1 }}}
\else
  \newcommand{\ian}[1]{}
  \newcommand{\liz}[1]{}
  \newcommand{\tak}[1]{}
  \newcommand{\takbefore}[1]{}
\fi

\def\ocpatches{\texttt{OC-Patches}}
\def\ocpatch{\texttt{OC-Patch}}
\def\modisAE{\texttt{OC-Patches\textsubscript{AE}}}
\def\modisHAC{\texttt{OC-Patches\textsubscript{HAC}}}
\def\modisHACtwo{\texttt{OC-Patches\textsubscript{HAC-2}}}
\def\modisHACthree{\texttt{OC-Patches\textsubscript{HAC-3}}}
\def\ocgridcell{\texttt{OC-Gridcell}}

\usepackage{changepage}
\usepackage{natbib}
\usepackage{siunitx}
\usepackage[colorlinks]{hyperref}

\usepackage{subfig}
\newcommand{\bftab}{\fontseries{b}\selectfont}
\usepackage{booktabs}

\title{AICCA: AI-driven Cloud Classification Atlas}

\author{%
    Takuya Kurihana\thanks{Globus Labs: \url{https://labs.globus.org/}} \\
    Dept. of Computer Science\\
    University of Chicago\\
    Chicago, IL 60637 \\
    \texttt{tkurihana@uchicago.edu} \\
\And
    Elisabeth J. Moyer\thanks{The Center for Robust Decision-making on Climate and Energy Policy (RDCEP): \url{http://www.rdcep.org/}} \\
    Dept. of the Geophysical Sciences\\
    University of Chicago\\
    Chicago, IL 60637 \\
    \texttt{moyer@uchicago.edu} \\
\And
    Ian T. Foster\footnotemark[1] \thanks{Argonne National laboratory, The Data Science and Learning Division: \url{https://www.anl.gov/dsl}} \\
    Dept. of Computer Science\\
    University of Chicago\\
    Chicago, IL 60637 \\
    \texttt{foster@uchicago.edu} \\
}

\begin{document}
\maketitle

\begin{abstract}
Clouds play an important role in the Earth’s energy budget, and their behavior is one of the largest uncertainties in future climate projections. Satellite observations should help in understanding cloud responses, but decades and petabytes of multispectral cloud imagery have to date received only limited use. This study describes a new analysis approach that reduces the dimensionality of satellite cloud observations by grouping them via a novel automated, unsupervised cloud classification technique based on a convolutional autoencoder, an artificial intelligence (AI) method good at identifying patterns in spatial data.
Our technique combines a rotation-invariant autoencoder and hierarchical agglomerative clustering to generate cloud clusters that capture meaningful distinctions among cloud textures, using only raw multispectral imagery as input. Cloud classes are therefore defined based on spectral properties and spatial textures without reliance on location, time/season, derived physical properties, or pre-designated class definitions. We use this approach to generate a unique new cloud dataset, the AI-driven cloud classification atlas (AICCA), which clusters 22 years of ocean images from the Moderate Resolution Imaging Spectroradiometer (MODIS) on NASA’s Aqua and Terra instruments---198 million patches, each roughly 100 km $\times$ 100 km (128 $\times$ 128 pixels)---into 42 AI-generated cloud classes, a number determined via a newly-developed stability protocol that we use to maximize richness of information while ensuring stable groupings of patches. 
AICCA thereby translates 801 TB of satellite images into  54.2 GB of class labels and cloud top and optical properties, a reduction by a factor of {15,000}. 
The 
42 AICCA classes produce meaningful spatio-temporal and physical distinctions and capture a greater variety of cloud types than do the nine International Satellite Cloud Climatology Project (ISCCP) categories---for example, multiple textures in the stratocumulus decks along the West coasts of North and South America. 
We conclude that our methodology has explanatory power, capturing 
regionally unique cloud classes and providing rich but tractable information for global analysis.
AICCA delivers the information from multi-spectral images in a compact form, enables data-driven diagnosis of patterns of cloud organization, provides insight into cloud evolution on timescales of hours to decades, and helps democratize climate research by facilitating access to core data.
\end{abstract}




\section{Introduction}

Over the past several decades, advancements in satellite-borne remote sensing instruments have produced petabytes of global multispectral imagery that capture cloud structure, size distributions, and radiative properties at a near-daily cadence. While understanding trends in cloud behavior is arguably the principal challenge in climate science, these enormous datasets are underutilized because climate scientists cannot in practice manually examine them to analyze spatial-temporal patterns. Instead, some kind of automated algorithm is needed to identify physically relevant cloud types. However, the diversity of cloud morphologies and textures, and their multi-scale properties, makes classifying them into meaningful groupings a difficult task.

Existing classification schemes are necessarily simplistic. The most standard classification, the ISCCP (International Satellite Cloud Climatology Project) schema, simply defines a grid of nine global classes based on low, medium, or high values of cloud altitude (cloud top pressure) and optical thickness~\cite{isccp1991,rossow1993comparison,rossow1999advances}. Because this classification is typically applied pixel by pixel, it cannot capture spatial structures and can produce an incoherent spatial distribution of cloud types in cloud imagery. The World Meteorological Organization’s International Cloud Atlas~\cite{wmo2017}, a more complex cloud classification framework, defines 28 different classes (of which 10 are considered ‘basic types’) with a complex coding procedure that depends on subjective judgments, such as whether a cloud has yet ``become fibrous or striated.'' The schema is subjective and difficult to automate, and furthermore does not capture the full diversity of important cloud types. For example, it does not distinguish between open- and closed-cell stratocumulus clouds, placing them both in ``stratocumulus,'' though the two have different circulation patterns, rain rates, and radiative effects~\cite{woodStCreview}.  Because the human eye serves as a sensitive tool for pattern classification, human observers can in principle group clouds into a larger set of types based on texture and shape as well as altitude and thickness. In practice, however, it has been difficult to devise a set of artificial cloud categories that encompass all cloud observations and can be applied consistently by human labelers.

These issues motivate the application of artificial intelligence (AI)-based algorithms for cloud classification. In the last several years, a number of studies have sought to develop AI-based cloud classification by using \textit{supervised learning}~\cite{Zhang2018CloudNetGC, Rasp2019CombiningCA, Zantedeschi2019CumuloAD,yuan2020applying, marais2020leveraging}. In these approaches, ML models are trained to classify cloud images based on a training set to which humans have assigned labels. However, the difficulty of generating meaningful and consistent labels is a constant problem, and supervised learning approaches tend to succeed best when used on limited datasets containing classic examples of well-known textures. For example, \citet{Rasp2019CombiningCA} classified just four particular patterns of stratocumulus defined and manually labeled by \citet{stevens2020sugar}. Supervised methods cannot discover unknown cloud types that may be relevant to climate change research.

To serve the needs of climate research {free from assumptions that may limit novel discoveries}, the more appropriate choice is \textit{unsupervised learning}, in which unknown patterns in data are learned without requiring predefined labels. The first demonstrations of unsupervised methods applied to cloud images were made in the 1990s~\cite{Visa1995NeuralNB, Tian1999ASO}. Even with the primitive neural networks then available, \citet{Tian1999ASO} showed that cloud images from the GOES-8 satellite could be sorted automatically into ten clusters that reproduced the ten ‘basic’ WMO classes with 65--75\% accuracy. In 2019, \citet{denby2019unsuper} and \citet{kurihana2019cloud} leveraged advances in deep neural network (DNN) methods to prototype unsupervised cloud classification algorithms that used convolutional neural networks (CNNs, DNNs with convolutional layers) and produced cloud classes from the resulting compact representations via hierarchical agglomerative clustering (HAC)~\cite{Johnson1967HAC}. Both works used only 12 classes and neither was rotation-invariant, but both successfully produced reasonable-seeming classifications---for \citet{denby2019unsuper}, from near-infrared images from the GOES satellite in the tropical Atlantic, and for \citet{kurihana2019cloud}, from global multispectral images from the Moderate Resolution Imaging Spectroradiometer (MODIS) instruments on NASA’s Aqua and Terra satellites).
\citet{kurihana2019cloud} were the first to use an autoencoder~\cite{Hinton2011TransformingA}, a class of unsupervised DNNs widely used for dimensionality reduction, for cloud classification, and \citet{kurihanaRICC21} extended the work by adding a more complex loss function to the autoencoder to produce rotation-invariant cloud clustering (RICC). \citet{kurihanaRICC21} also developed a formal evaluation protocol to ensure that the resulting cloud classes were physically meaningful.

The work described here builds on these previous results to generate a standardized science product: an AI-driven Cloud Classification Atlas (AICCA) of global-scale unsupervised classification of MODIS satellite imagery into 42 cloud classes. 
We first describe and apply the protocol that we have developed to determine this optimal number of clusters when applying RICC to the MODIS {dataset}. ({The first author calculated this number before being informed of its occurrence in an unrelated }context~\cite{Hitchhikers}.) We demonstrate that the resulting classes are coherent geographically, temporally, and in altitude-optical depth space. 
Finally, we describe a workflow that allows us to apply the RICC\textsubscript{42} algorithm to the full two decades of MODIS imagery to provide a publicly available dataset. The result is an automated, unsupervised classification process that discovers classes based on both cloud morphology and physical properties to yield unbiased cloud classes free from artificial assumptions that capture the diversity of global cloud types. AICCA is intended to support studies of the response of clouds to forcing on timescales from hours to decades and to allow data-driven diagnosis of cloud organization and behavior and their evolution over time as CO$_2$ and temperatures increase.

We describe this dataset as follows: Section~\ref{sec:materials}  describes the MODIS imagery, information used, and structure of output data. Section~\ref{sec:method}  describes the algorithm used for classification, including the training procedure on one million randomly selected ocean-cloud patches (Section~\ref{sec:RICC}). Section~\ref{sec:stability} evaluates the stability of the clustering step, and  Section~\ref{sec:results} describes the characteristics of the resulting cloud clusters: their distribution geographically, seasonally, and in altitude-optical depth space.

\section{AICCA: Data and Outputs}\label{sec:materials}

The dataset described in this article,
AICCA\textsubscript{42} (or simply AICCA), provides AI-generated cloud class labels for all 128 $\times$ 128 pixels ($\sim$100 km by 100 km) ocean cloud \textit{patches} sampled by MODIS instruments over their 22 years of operation.
(An ocean cloud patch is defined as a patch with only ocean pixels and at least 30\% cloud pixels.)
The cloud labels are generated by the rotation-invariant cloud clustering (RICC) method of \citet{kurihanaRICC21}.
In general, clusters produced by RICC may vary according to (1) the patches used to train RICC, (2) the number of clusters chosen, and (3) the patches to which the trained RICC is applied to generate centroids.
We therefore define AICCA\textsubscript{42} as the dataset produced by training RICC on a subset of the data described in Section~\ref{sec:modis}, clustered into 42 classes with a set of reference centroids based on \modisHAC, as defined in Section~\ref{sec:seasonal_clusters}.

The labeled output is provided in two ways: per patch, which provides the finest granularity of labels and associated physical properties, and resampled to $1^{\circ} \times 1^{\circ}$ grid cells, which supplies information in a daily global grid format that is familiar to climate~scientists.

\subsection{MODIS Data}\label{sec:modis}

The MODIS instruments hosted on NASA’s Aqua and Terra satellites have been collecting visible to mid-infrared radiance data in 36 spectral bands from 2002 (Aqua)~\cite{myd02} and 2000 (Terra)~\cite{mod02} through 2021.  
The instruments collect data over an approximately {2330} km by {2030} km \textit{swath} every five minutes, with a spatial resolution of 1~km. 
AICCA is based on the MODIS Level 1B calibrated radiance product (MOD02). \footnote{{The first author calculated this number before being informed of its occurrence in an unrelated context~\cite{Hitchhikers}.}}. 
We limit the dataset to the six spectral bands most relevant for derivation of physical properties: bands 6, 7, and 20 relate to cloud optical properties, and bands 28, 29, and 31 relate to the separation of high and low clouds and the detection of the cloud phase. 
For the Aqua instrument, we use band 5 as an alternative to band 6 due to a known stripe noise issue in
Aqua band 6~\cite{rakwatin2007stripe}. (See also \citet{kurihanaRICC21} for more details.)
The total number of swath images per band is 
(12 swath/h) $\times$ (12 h/day) $\times$ (365 day/year) $\times$ (20 + 22 years, for Aqua and Terra, respectively) $\approx$ 2.2 million.

MODIS multispectral data are processed by NASA to yield a variety of derived products, several of which we employ for post-processing or analysis. We take latitude and longitude from the MOD03 geolocation fields to regrid the AICCA patches, and use selected derived physical properties from the MOD06 product to evaluate the cloud classes: four physical parameters related to cloud optical properties and cloud top properties. 
Note that we employ the MOD06 variables only as a diagnostic, to evaluate associations between AICCA clusters and cloud physical properties. 
They are not included in our RICC training data, which are thus free from any assumptions made by the producers of MOD06 variables.
The data used in generating AICCA, listed in Table~\ref{tab:products}, have an aggregate size of 801 terabytes. All MODIS products are accessible via the NASA Level-1 and Atmosphere Archive and Distribution System (LAADS), grouped into per-swath files.


\begin{table}[!ht]
\caption{MODIS products used to create the AICCA dataset. 
Each product name \textit{MOD0X} in the first column refers to both the Aqua (MYD0X) and Terra (MOD0X) products.
Source: NASA Earthdata.}\label{tab:products}
\begin{adjustwidth}{-0cm}{0cm}

	\newcolumntype{C}{>{\centering\arraybackslash}X}
		\begin{tabularx}{0cm}{m{1.5cm}<{\raggedright}m{6.5cm}<{\raggedright}m{1cm}<{\centering}m{4.5cm}<{\raggedright}Cm{1.5cm}<{\raggedright}}
\cline{1-6}
 \textbf{Product} & \textbf{Description} &\textbf{Band}& \textbf{Primary Use } & &\textbf{Process} \\ 
\cline{1-6}
 MOD02   & Shortwave infrared (1.230--1.250 $\upmu$m)         & 5 & Land/cloud/aerosol properties & \rdelim\}{7}{0.1cm}& \multirow{7}{1cm}{Section~\ref{sec:stage1}} \\
         & Shortwave infrared (1.628--1.652 $\upmu$m)         & 6 & Land/cloud/aerosol properties &&   \\
         & Shortwave infrared (2.105--2.155 $\upmu$m)         & 7 & Land/cloud/aerosol properties &&  \\
         & Longwave thermal infrared (3.660–3.840 $\upmu$m)  & 20& Surface/cloud temperature  && \\
         & Longwave thermal infrared (7.175–7.475 $\upmu$m)  & 28& Cirrus clouds water vapor   &&\\
         & Longwave thermal infrared (8.400–8.700 $\upmu$m)  & 29& Cloud properties&& \\
         & Longwave thermal infrared (10.780–11.280 $\upmu$m)& 31& Surface/cloud temperature   &&\\ 
         \cline{1-4}
  MOD03  & Geolocation fields     &  & Latitude and Longitude  & \rdelim\}{3}{0.1cm}& \multirow{3}{1cm}{Section~\ref{sec:stage1}}\\ 
  \cline{1-4}
  MOD06  & Cloud mask             &  & Cloud pixel detection   &&\\
         & Land/Water           &  & Background detection   &&\\
         & Cloud optical thickness&  & Thickness of cloud      & \rdelim\}{4}{*}& \multirow{4}{0.1cm}{Section~\ref{sec:evaluation}} \\
         & Cloud top pressure     &  & Pressure at cloud top  & &\\
         & Cloud phase infrared   &  & Cloud particle phase    &&\\
         & Cloud effective radius &  & Radius of cloud droplet& &\\ 
         \cline{1-6}
\end{tabularx}
\end{adjustwidth}
\end{table}

\subsection{AICCA Patch-Level Data}\label{sec:patch}

The AICCA dataset uses all patches from Aqua and Terra MODIS image data during~2000--2021, subject to the constraints that they (1) are
disjoint in space and/or time; (2) include no non-ocean pixels, and 3) each includes at least 30\% cloud pixels.
The resulting set comprises about {198,676,800} individual 128 $\times$ 128 pixel ($\sim$100 km by 100 km) ocean-cloud patches, for each of which 
AICCA\textsubscript{42} provides the following information (and see Table~\ref{tab:patch_values}): 
\begin{itemize}
    \item
    \texttt{Source} is either Aqua or Terra;
    \item 
    \texttt{Swath}, \texttt{Location}, and \texttt{Timestamp} locate the patch in time and space;
    \item 
    \texttt{{Training}} indicates whether the patch was used for training;
    \item 
    \texttt{{Label}} is an integer in the range 1{..}42, generated by the rotation-invariant cloud clustering system configured for 42 clusters, RICC\textsubscript{42} {(see Section~\ref{sec:stability} for the stability protocol used to select this number of clusters});
    \item 
    \texttt{COT\_patch}, \texttt{CTP\_patch}, and \texttt{CER\_patch},
    the mean and standard deviation, across all pixels in the patch, for three MOD06 physical values: cloud optical thickness (COT), cloud top pressure (CTP), and cloud effective radius (CER); and
    \item 
    \texttt{CPI\_patch}, 
    cloud phase information (CPI), four numbers representing the number of the 128 $\times$ 128 pixels in the patch that are estimated as clear-sky, liquid, ice, or undefined,~respectively.
\end{itemize}

The resulting 146~bytes per patch represents a {16,159} $\times$ reduction in size
relative to the raw multispectral imagery.

The additional information shown in Table~\ref{tab:patch_values}
to assist users in understanding individual patches
is extracted from MOD06 by using the patch's geolocation index and timestamp (Location and Timestamp in Table~\ref{tab:patch_values}) to locate the patch's data in the appropriate MOD06 file. 
These mean values summarize the patch's average physical characteristics; 
the standard deviations provide some indication as to the existence of multiple clouds (especially low- and high-altitude clouds).
We do not use the MOD06 multilayered cloud~flag.

Output is provided as NetCDF~\cite{rew1990netcdf} files that combine patches from each MODIS swath into a single file.
While AICCA contains no raw satellite data, it includes for each patch an identifier for the source MODIS swath and a geolocation index; thus, users can easily link AICCA results with the original MOD02 satellite imagery and other MODIS products. 
The complete \ocpatches{} set contains around (20 $+$ 22 years) $\times$ (365 days/year) $\times$ ({26,000} patches) $\times  $146~B $\approx$ 54.2 gigabytes.

\begin{table}[!ht]
    \caption{{Information} provided in AICCA  for each 128$\times$128 pixel ocean-cloud patch: metadata that locate the patch in space and time, and indicate whether the patch was used to train RICC; a cloud class label computed by RICC; and a set of diagnostic quantities obtained by aggregating MODIS data over all pixels in the patch. A quotation mark indicates a repetition. 
    }
    \label{tab:patch_values}
\newcolumntype{C}{>{\centering\arraybackslash}X}
\begin{tabularx}{\textwidth}{m{1.7cm}<{\centering}m{8.6cm}<{\centering}m{0.8cm}<{\centering}m{1cm}<{\centering}}
\hline
    \textbf{Variables}  & \textbf{Description} & \textbf{Values}  & \textbf{Type} \\
    \hline
    Swath & Identifier for source MODIS swath & 1 &  float32 \\
    Location & Geolocation index for the upper left corner of patch & 2 &  float32 \\    
    Timestamp & Time of observation & 1 &  float32 \\  
    Training & Whether patch used for training & 1 &  binary \\    
    \hline 
    Label & Class label assigned by RICC: integer in range 1{..}$k^{\ast}$& 1 &  int32 \\
    \hline
    COT\_patch & Mean and standard deviation of pixel values in patch & 2 & float32 \\
    CTP\_patch & {"} & {"} & {"} \\
    CER\_patch &  " & " & " \\ 
    CPI\_patch & Number of pixels in patch in \{clear-sky, liquid, ice, undefined\} & 4 & int32 \\ 
    \hline
     \end{tabularx}
\end{table}




\subsection{AICCA Grid Cell-Level Data}\label{sec:aicca_gridcell}

In addition to providing per-patch data, we follow common practice in climate datasets by also providing data organized on a per-latitude/longitude grid cell basis.
The second element of the AICCA\textsubscript{42 } dataset spatially aggregates the patch-level class label and diagnostic values at a resolution of $1^{\circ} \times 1^{\circ}$, a total of 181 $\times$ 360 \textit{{grid cells}} over the globe.
For each grid cell, 
AICCA\textsubscript{42} provides the information listed in Table~\ref{tab:cell_values}, a total of 32 bytes: 
\begin{itemize}
    \item
    \texttt{{Source}} is either Aqua or Terra;
    \item 
    \texttt{{Cell}} gives a latitude and longitude for the grid cell;
    \item
    \texttt{{Timestamp}} locates the grid cell in time;
    \item 
    \texttt{Label\_1deg} represents the most frequent class label in the grid cell (an integer in the range 1..42); and 
    \item 
    \texttt{COT\_1deg}, \texttt{CTP\_1deg}, \texttt{CER\_1deg}, and \texttt{CPI\_1deg} aggregate values for four diagnostic variables, as described in Section~\ref{sec:aicca_gridcell}. 
\end{itemize}

The aggregation process uses values from individual days from the Aqua and Terra satellites,
a reasonable choice since the 
swaths taken by each satellite's MODIS instrument generally do not overlap in a daily period. 
Since a single 2330 km by 2030 km MODIS swath extends across multiple 1 degree by 1 degree grid cells, we extract the latitude and longitude at the center of each \ocpatch{} by using MOD03,
and aggregate the information listed in Table~\ref{tab:patch_values} to each 1$^{\circ}$ $\times$ 1$^{\circ}$ grid cell (i.e., the area extending from $-0.5^{\circ}$ to $+0.5^{\circ}$ from the grid cell center). To assign a class label to each grid cell on each day, we use the class of the single ocean-cloud patch with the largest overlap with the grid cell. 
To provide physical properties for each grid cell, we implement one simplification to reduce the use of computing memory: instead of averaging pixel values within each grid cell, we identify \textit{{all}} ocean-cloud patches that overlap with the cell, and simply average those patches' mean COT, CTP, and CER values.
To assign a cloud particle phase (clear--sky, liquid, ice, or undefined), we use the most frequent phase in the overlapping patches.
Grid cells with no clouds are labeled as a missing value.

In some cases, especially at high latitudes, swaths may overlap within a single day. When this occurs, patches with different timestamps will overlap a given grid cell on the same day. In these cases, we discard one timestamp, to avoid inconsistent values between grid cells. That is, when accumulating the most frequent label and aggregating values on the overlapping cell, 
we use only those patches with a timestamp close to that of the neighboring grid cells.
This neighboring selection mitigates the problem of inconsistent values between nearly grid cells due only to timing.
Finally, we accumulate the aggregated grid-cell values to create the daily files.
Given the MODIS orbital coverage, the complete \ocgridcell{} set contains around (20 $+$ 22 years) $\times$ (365 days/year) $\times$ ({65,160} grid cells) $\times$ 32~B $\approx$ 29.8 gigabytes.

\begin{table}[!ht]
    \caption{AICCA information for each $1^{\circ} \times 1^{\circ}$ grid cell: a cloud class label computed by RICC and diagnostic quantities obtained by aggregating MODIS data over all patch pixels for that grid cell.
    }
    \label{tab:cell_values}
  \newcolumntype{C}{>{\centering\arraybackslash}X}
\begin{tabularx}{\textwidth}{Cm{6cm}<{\centering}CC}
\hline
    \textbf{Variables}  & \textbf{Description} & \textbf{Values}  & \textbf{Type} \\
    \hline
    Cell & (lat, long) for grid cell & 2 &  float32 \\    
    Timestamp & Time of observation & 1 &  float32 \\  
\hline
    Label & Most frequent class label in grid cell & 1 &  int32 \\
    \hline
    COT\_1deg & Mean of pixel values in grid cell & 1 & float32 \\
    CTP\_1deg & {"} & {"} & {"} \\
    CER\_1deg &  " & " & " \\ 
    CPI\_1deg & Most frequent particle phase in grid cell & 1 & int32 \\ 
    \hline
     \end{tabularx}
\end{table}

\vspace{-12pt}

\section{Constructing AICCA}\label{sec:method}
\begin{figure}[!ht]
    \begin{adjustwidth}{-0cm}{0cm}
    \centering
    \includegraphics[width=18.4cm,trim=1mm 2mm 0mm 2mm,clip]{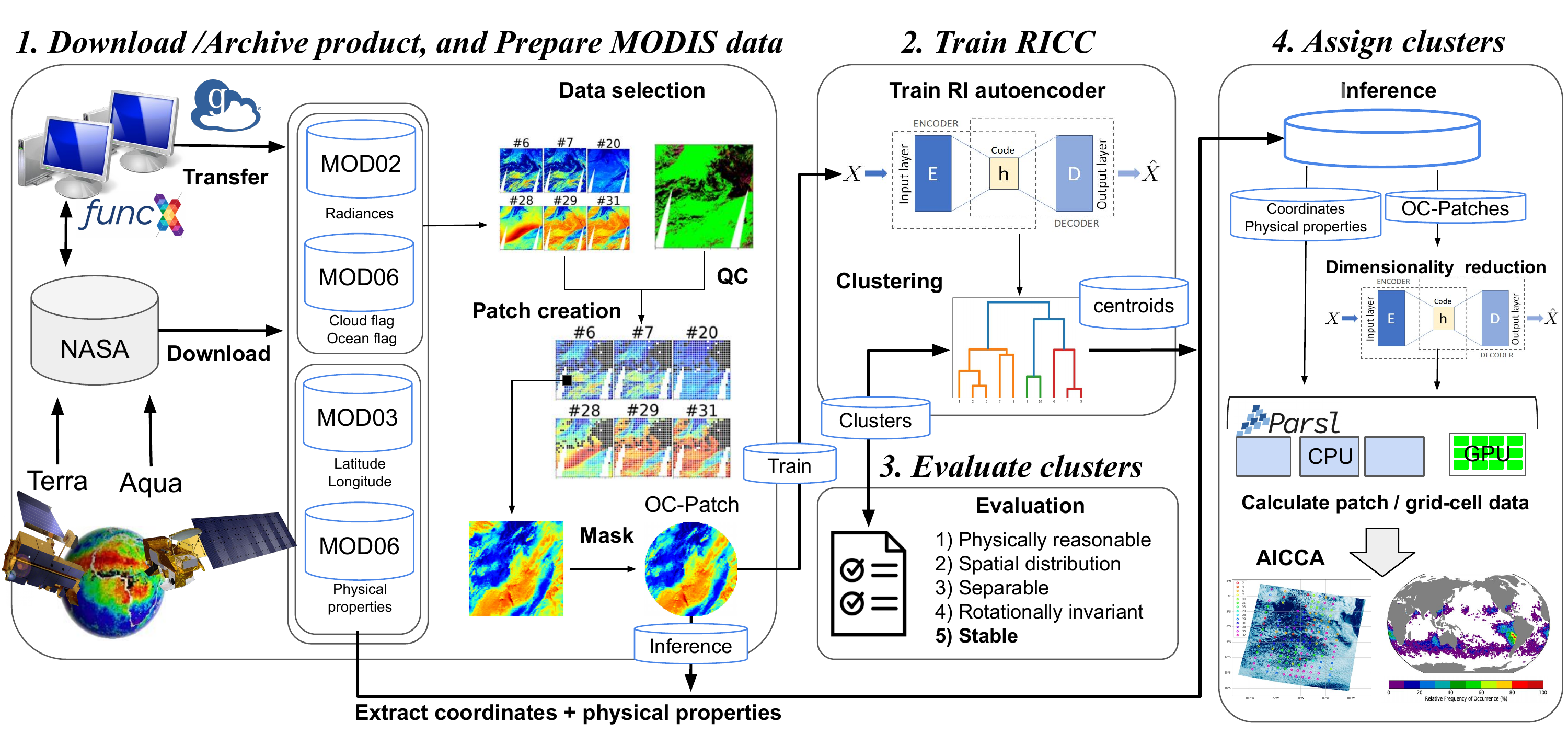}
    \end{adjustwidth}
    \caption{The AICCA production workflow comprises four principal stages. 
    \textbf{{(1) Download/Archive and Prepare MODIS data}}: Download calibrated and retrieved MODIS products from the NASA Level-1 and Atmosphere Archive and Distribution System (LAADS), using FuncX and Globus for rapid and reliable retrieval of {801} terabytes of three different MODIS products between 2000--2021. 
    Store downloaded data at Argonne National Laboratory.
    Select six near-infrared to thermal bands related to clouds and subdivide each swath into non-overlapping 128 $\times$ 128 pixel patches by six bands. 
    Select patches with >30\% cloud pixels over ocean regions, and apply a circular mask for optimal training of our rotation-invariant autoencoder, yielding \ocpatches{}.
    \textbf{{(2)~Train RICC}}: Train an autoencoder on 1 M randomly selected patches to generate latent representations, and cluster those latent representations to determine cluster centroids~\cite{kurihanaRICC21}.  
    \textbf{(3)~Evaluate clusters}: Apply five protocols to evaluate whether the clusters produced are meaningful and useful.  
    \textbf{(4) Assign clusters}: Use trained autoencoder and centroids to assign cloud labels to unseen data. We use the Parsl parallel Python library to scale the inference process to hundreds of CPU nodes plus a single GPU, and to generate the AICCA dataset in NetCDF format. 
    We then calculate physical properties and other metadata information for each patch and for each 1$^{\circ}$ $\times$ 1$^{\circ}$ grid cell.
    }\label{fig:aicca_workflow}
\end{figure}

The AICCA production workflow, shown in Figure~\ref{fig:aicca_workflow}, consists of four principal stages: (1) download, archive, and prepare MODIS satellite data; (2) train the RICC unsupervised learning algorithm, and cluster cloud patterns and textures; (3) evaluate the reasonableness of the resulting clusters and determine an optimal cluster number; and (4) assign clusters produced by RICC to other MODIS data unseen during RICC training.
We describe each stage in turn.  The RICC code and Jupyter notebook~\cite{kluyver2016jupyter} used in the analysis are available online~\cite{RICCcode}, and the trained RI autoencoder used for this study is archived at the Data and Learning Hub for science (DLHub)~\cite{chardDLhub19},
a scalable and low-latency model repository to share and publish machine learning models to facilitate reuse and reproduction.

\subsection{Stage 1: Download, Archive, and Prepare MODIS Data}\label{sec:stage1}

\textit{{Download and archive.}} As noted in Section~\ref{sec:modis}, we use subsets of three MODIS products in this work, a total of 801 terabytes for 2000--2021.
In order to employ high-performance computing resources at Argonne National Laboratory for AI model training and inference, we copied all files to Argonne storage. 
Transferring the files from NASA archives is rapid for the subset that are accessible on a Globus endpoint at the NASA Center for Climate Simulation, which can be transferred via the automated Globus transfer system~\cite{chard2014efficient}.
The remaining files were transferred from NASA LAADS via the more labor-intensive option of wget commands, which we accelerated by using the funcX~\cite{chard2020funcx} 
distributed function-as-a-service platform to trigger concurrent downloads on multiple~machines.

{{\textit{{Prepare}.}} The next step involves preparing the {\textit{{patches}}} used for ML model training and inference. 
~We extract from each swath multiple 128 pixel by 128 pixel (roughly 100 km $\times$ 100~km)} non-overlapping patches, for a total of $\sim$331 million patches. We then eliminate those patches that include any non-ocean pixels as indicated by the MOD06 land/water indicator, since, in these cases, radiances depend in part on underlying topography and reflectance. (Note that even ocean-only pixels may involve surface-related artifacts in cases when the ocean is covered in sea ice.)
We also eliminate those with less than 30\% cloud pixels, as indicated by the MOD06 cloud mask.
The result is a set of {198,676,800} \textit{{ocean-cloud patches}}, which we refer to in the following as \ocpatches{}.
For each ocean-cloud patch, we take from the MOD02 product six bands (out of 36 total) for use in training and testing the rotation-invariant (RI) autoencoder. We also extract the MOD04 and MOD06 data used for location and cluster evaluation, as described in Section~\ref{sec:materials}.
For an in-depth discussion of data selection, see~\citet{kurihanaRICC21}.

We also construct a training set \modisAE{} by selecting one million patches at random from the entirety of \ocpatches{}.
Because we do not expect our unsupervised RI autoencoder to be robust to the MODIS data used for training, 
we collect the 1M patches that they are not overly imbalanced among seasons or locations. 

\subsection{Stage 2: Train the RICC Autoencoder and Cluster Cloud Patterns}\label{sec:RICC}
In this stage, we first train the RI autoencoder and then define cloud categories by clustering the compact latent representations produced by the trained autoencoder. 

\textit{{Train RICC}.} The goal of training is to produce an RI autoencoder capable of generating latent representations (a lower-dimensional embedding as the intermediate layer of the autoencoder) that explicitly capture the variety of input textures among ocean clouds and also map to differences in physical properties. 
We introduce general principles briefly here; see \citet{kurihanaRICC21} for 
further details of the RI autoencoder architecture and training~protocol. 

An autoencoder~\cite{Hinton94AEscience,Hinton2011TransformingA} is a widely used unsupervised learning method that leverages dimensionality reduction as a preprocessing tool prior to image processing tasks such as clustering, regression, anomaly detection, and inpainting. 
An autoencoder comprises an encoder, used to map input images into a compact lower-dimensional latent representation, followed by a decoder, used to map that representation to output images. During training, a loss function minimizes the difference between input and output. The resulting latent representation in the trained autoencoder both (1) retains only relevant features for the target application in input images, and (2) maps images that are similar (from the perspective of the target application) to nearby locations in latent space.

\begin{figure}[!ht]
    \begin{minipage}{.49\columnwidth}
        \subfloat[]{\includegraphics[clip, width=0.99\textwidth]{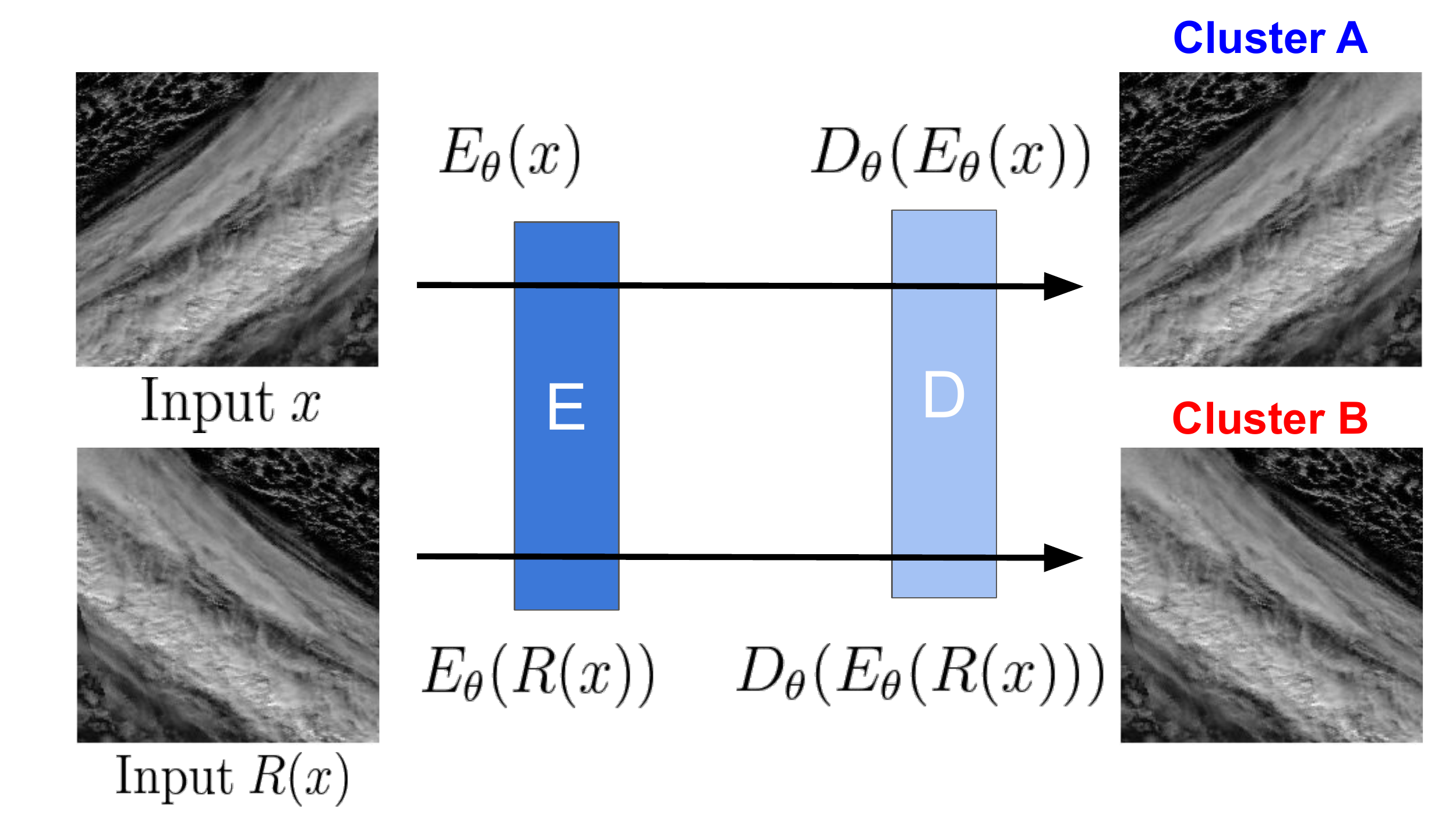}
        \label{fig:conceptOrdAE}}
    \end{minipage}
    \begin{minipage}{.49\columnwidth}
        \subfloat[]{\includegraphics[clip, width=0.99\textwidth]{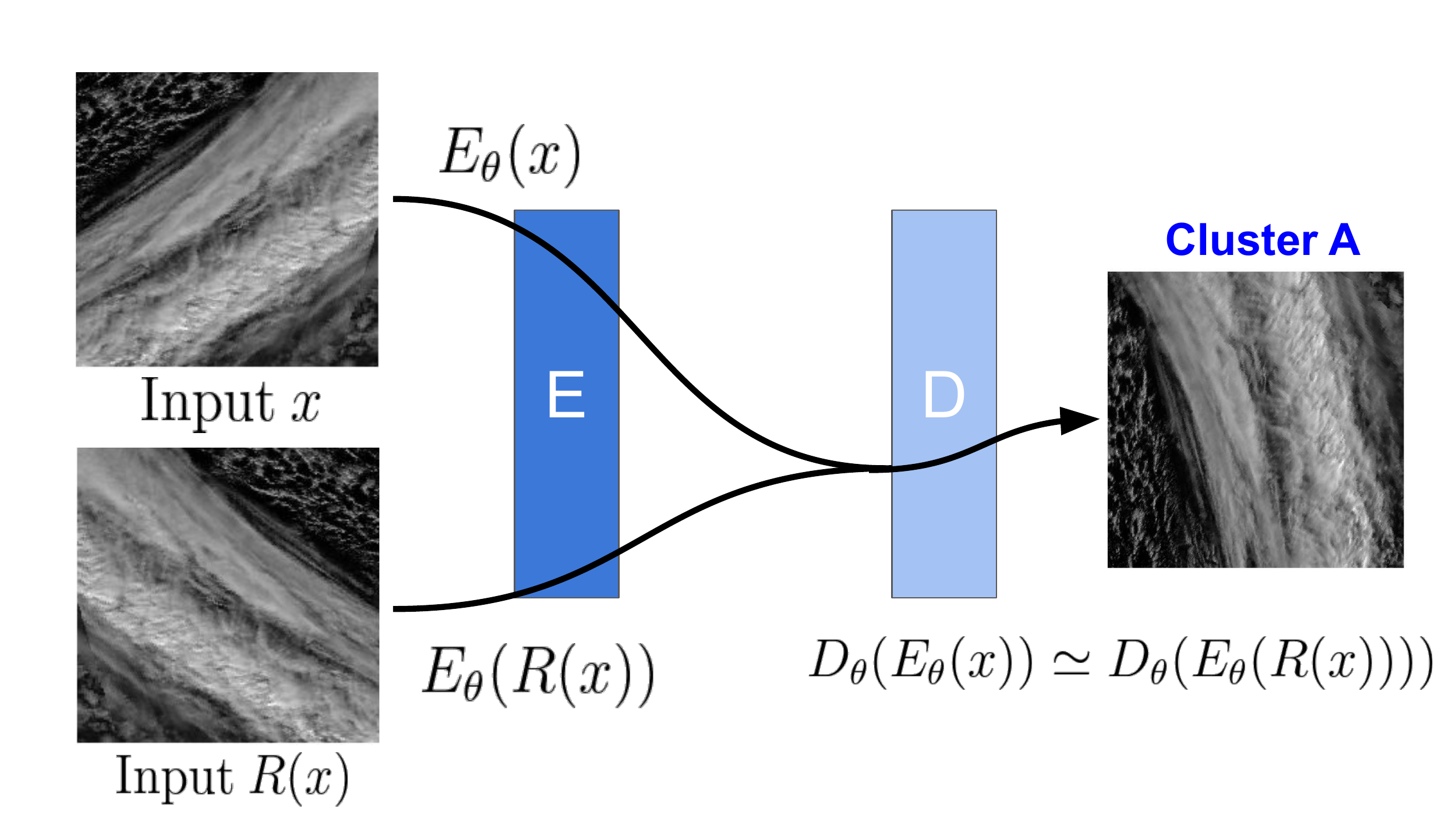}
        \label{fig:conceptRI}}
    \end{minipage}
    \caption{{Illustration of the learning }process when training (\textbf{a}) a conventional autoencoder with Equation~(\ref{standardloss}) vs.\ (\textbf{b}) a rotation-invariant autoencoder with Equation~(\ref{eq:ri_loss}). 
    Because a conventional autoencoder reflects orientation in the latent representation, two input images that are identical in texture but different in orientation are assigned to different clusters, A and B.
    The rotation-invariant autoencoder produces a latent representation that is agnostic to orientation, allowing clustering to group both together.
 }\label{fig:concept-AEs}
\end{figure}

The loss function minimizes the difference between an original and a restored image based on a distance metric during autoencoder training. 
The most commonly used metric is a simple ${\ell}^2$ distance between the autoencoder's input and output:
\begin{equation} \label{standardloss}
    L(\boldsymbol{\theta}) = \sum_{x \in S} || x - D_{\theta}(E_{\theta}(x))  ||^2_2 ,
\end{equation}
where $S$ is a set of training inputs; 
$\boldsymbol{\theta}$ is the encoder and decoder parameters, for which values are to be set via training; and
$x$ and $D_{\theta}(E_{\theta}(x))$ are an input in $S$ and its output (i.e., the restored version of $x$), respectively.
However, optimizing with Equation~(\ref{standardloss}) is inadequate for our purposes because it tends to generate different representations for an image $x$ and the rotated image $R(x)$, as shown in Figure~\ref{fig:concept-AEs}, with the result that the two images end up in different clusters. 
Since any particular physically driven cloud pattern can occur in different orientations,
we want an autoencoder that assigns cloud types to images consistently, regardless of orientation. 
Other ML techniques that combine dimensionality reduction with clustering algorithms have not addressed the issue of rotation--invariance within their training process. For example, while non-negative matrix factorization (NMF)~\cite{lee1999learning} can approximate input data into a low-dimensional matrix---i.e., 
produce a dimensionally reduced representation similar to an autoencoder---that can be used for clustering, 
applications of NMF are not invariant to image orientation.

We have addressed this problem in prior work by defining a rotation-invariant loss function~\cite{kurihanaRICC21} that generates similar latent representations, agnostic to orientation, for similar morphological clouds  (Figure~\ref{fig:concept-AEs}b). 
This RI autoencoder, motivated by 
the shifted transform invariant autoencoder of Matsuo et al.~\cite{Matsuo2017TransformIA}, uses
a loss function $L$ that combines both a rotation-invariant loss, $L_{\mathrm{inv}}$, to learn the rotation invariance needed to map different orientations of identical input images into a uniform orientation, and a restoration loss, $L_{\mathrm{res}}$, to learn the spatial structure needed to restore structural patterns in inputs with high fidelity. The two loss terms are combined as follows, with values for the scalar weights $\lambda_{\text{inv}}$ and $\lambda_{\text{res}}$ chosen as described below:
\begin{equation}\label{eq:ri_loss}
    L  = \lambda_{ \text{inv}} L_{\text{inv}} + \lambda_{ \text{res}} L_{\text{res}},
\end{equation}


The rotation-invariant loss function $L_{ \text{inv}}$ computes, for each image in a minibatch, the difference between the restored original and the 72 images obtained by applying a set ${\cal R}$ of 72 scalar rotation operators, each of which rotates an input by a different number of degrees in the set \{0, 5, ..., 355\}:
\begin{equation}\label{tinv}
     L_{\mathrm{inv}}(\boldsymbol{\theta}) = \frac{1}{N}\sum_{x \in S}\sum_{R\in {\cal R}} {|| D_{\theta}(E_{\theta}(x)) - D_{\theta}(E_{\theta}(R(x)))||^2_2}.
\end{equation}

Thus, minimizing Equation~(\ref{tinv}) yields values for $\boldsymbol{\theta}$ that produce similar latent representations for an image, regardless of its orientation.


The restoration loss, $L_{\text{res}}(\boldsymbol{\theta})$, learns the spatial substructure in images by 
computing the sum of minimum differences over the minibatch:
\begin{equation}\label{tres}
    L_{\text{res}}(\boldsymbol{\theta}) = \sum_{x \in S} \displaystyle \min_{R \in {\cal R}} {|| R(x) - D_{\theta}(E_{\theta}(x)) ||^2_2} .
\end{equation}

Thus, minimizing Equation~(\ref{tres}) results in values for $\boldsymbol{\theta}$ that preserve spatial structure in~inputs.

Our RI autoencoder training protocol~\cite{kurihanaRICC21}, which sweeps over $\left(\lambda_{ \text{inv}},\lambda_{ \text{res}} \right)$ values, identifies $\left(\lambda_{ \text{inv}},\lambda_{ \text{res}} \right) = \left( 32, 80\right)$ as the coefficients for the two loss terms
that best balance the transform-invariant and restoration loss terms. 
{We note that the specific values of the two coefficients, not just their relative values, matter. For example,
the values $\left(\lambda_{ \text{inv}},\lambda_{ \text{res}} \right) = \left( 32, 80\right)$ give better results than $\left(\lambda_{ \text{inv}},\lambda_{ \text{res}} \right) = \left( 3.2, 8.0\right)$.}

The neural network architecture is the other factor needed to achieve rotation invariance: 
Following the heuristic approach of deep convolutional neural networks, we designed an encoder and decoder that stack five blocks of convolutions, each with three convolutional layers activated by leaky ReLU~\cite{Nair2010RectifiedLU}, and with batch normalization~\cite{Ioffe2015BatchNA} applied at the final convolutional layer in each block before activation.
We train our RI autoencoder on our one million training patches for 100 epochs by using stochastic gradient descent with a learning rate of 10$^{-2}$ on 32 NVIDIA V100 GPUs in the Argonne National Laboratory ThetaGPU cluster.

\textit{Cluster Cloud Patterns.} Once we have applied the trained autoencoder to a set of patches to obtain latent representations, we can then cluster those latent representations to identify the centroids that will define our cloud clusters. 
We use hierarchical agglomerative clustering (HAC)~\cite{Johnson1967HAC} for this purpose, and select Ward's method~\cite{Ward1963HierarchicalGT} for the linkage metric, so that 
HAC minimizes the variance of square distances as it merges clusters from bottom to top. 
We have shown in previous work~\cite{Jenkins2019DevelopingUL} that HAC clustering results outperform those obtained with other common clustering algorithms.

Given $N$ data points, a naive HAC approach requires $\mathcal{O}(N^2)$ memory to store the distance matrix used when calculating the linkage metric to construct the tree structure~\cite{moertini2018big}---which would be impractical for the one million patches in \modisAE{}.
Thus, we use a smaller set of patches, \modisHAC{}, comprising {74911} ocean-cloud patches from the year 2003 (the first year in which both Terra and Aqua satellites ran for the entire year concurrently) for the clustering phase.
We apply our trained encoder to compute latent representations for each patch in \modisHAC{} and then run HAC to group those latent representations into $k^{\ast}$ clusters, in the process identifying $k^{\ast}$ cluster centroids and assigning each patch in \modisHAC{} a cluster label, 1..$k^{\ast}$.
The sequential scikit-learn~\cite{varoquaux2015scikit} implementation of HAC that we use in this work takes around 10 hours to cluster the {74911} \modisHAC{} patches on a single core.
While we could use a parallelizable HAC algorithm~\cite{jin2015scalable,sumengen2021scaling,monath2021scalable} to increase the quantity of data clustered, this would not address the intrinsic limitation of our clustering process given the {801} terabytes of MODIS data.

\subsection{Stage 3: Evaluate Clusters Generated by RICC}\label{sec:evaluation}

A challenge when employing unsupervised learning is to determine how to evaluate results.
While a supervised classification problem involves a perfect ground truth against which to output can be compared, an unsupervised learning system produces outputs whose utility must be more creatively evaluated. 
Therefore, we 
defined in previous work a series of evaluation protocols to determine whether the cloud classes derived from a set of cloud images are meaningful and useful~\cite{kurihanaRICC21}. 
We seek cloud clusters that: (1) are {\emph{{physically reasonable}}} (i.e., embody scientifically relevant distinctions); (2) capture information on {\emph{{spatial distributions}}}, such as textures, rather than only mean properties; (3) are {\emph{{separable}}} (i.e., are cohesive, and separated from other clusters, in latent space); (4) are {\emph{{rotationally invariant}}} (i.e., insensitive to image orientation); and (5) are {\emph{{stable}}} (i.e., produce similar or identical clusters when different subsets of the data are used).
We summarize in Table~\ref{tab:fiveprotocols} these criteria and the quantitative and qualitative tests that we have developed to validate them.

\begin{table}[!ht]
\caption{Our five evaluation criteria protocol, as described in~\citet{kurihanaRICC21}, and protocols for meeting them. In that work, we used the first four criteria to demonstrate that our quantitative and qualitative evaluation protocols can distinguish useful from non-useful autoencoders, even when common ML metrics such as ${\ell}^2$ loss show insignificant differences. In the current work, we describe a protocol to ensure meeting the last criterion, stability.
}\label{tab:fiveprotocols}
\begin{adjustwidth}{-0cm}{0cm}
		\begin{tabularx}{-1cm}{  p{3.2cm}  p{4.7cm}  p{8.8cm} } 
		\hline
\cline{1-3}
\textbf{Criterion} & \textbf{Test} &  \textbf{Requirement} \\ 
\cline{1-3}
Physically reasonable
    & Cloud physics & Non-random distribution; median inter-cluster correlation $<$ 0.6  \\ 
    \hline
   
\multirow{3}{*}{Spatial distribution}
    & Spatial coherence & Spatially coherent clusters \\ 
    \cline{2-3}  
    & Smoothing  & Low adjusted mutual information (AMI) score  \\
    \cline{2-3} 
    & Scrambling & Low AMI score  \\  
    \hline
Separable   & Separable clusters  & No crowding structure \\ 
\hline 
Rotationally invariant & Multi-cluster & AMI score closer to {1.0} \\ 
\hline
\multirow{4}{*}{Stable}
    & Significance of cluster stability & Ratio of Rand Index $G/R \geq$ 1.01 \\ 
    \cline{2-3}  
    & Similarity of clusterings & Higher Adjusted Rand Index (ARI)  \\  
    \cline{2-3} 
    & Similarity of intra-cluster textures & Lower weighted average mean square distance  \\ 
    \cline{2-3} 
    & Clusters capture seasonal cycle & Minimal seasonal texture difference \\
 \cline{1-3}
\end{tabularx}
\end{adjustwidth}
\end{table}

In our previous work~\cite{kurihanaRICC21}, we showed that an analysis using RICC to separate cloud images into 12 clusters satisfies the first four of these criteria.
In this work, we describe how we evaluate the last criterion, {\it{{stability}}}. 
Specifically, we evaluate the extent to which RICC clusters cloud textures and physical properties in a way that is stable 
against variations in the specific cloud patches considered, and that groups homogeneous textures within each cluster.
We describe this process in Section~\ref{sec:stability} in the context of how we estimate the \emph{optimal} number of clusters for this dataset when maximizing stability and similarity in clustering. For the remaining criteria,
the clusters necessarily remain {rotationally invariant},  
and we present in Section~\ref{sec:results}  results further validating that the algorithm, when applied to a global dataset, produces clusters that show physically reasonable distinctions, are spatially coherent, and involve distinct textures (i.e., learn spatial information).



\subsection{Stage 4: Assign Cluster Labels to Patches}\label{sec:inference}

We have so far trained our RI autoencoder on the 1 million patches in \modisAE{} and applied HAC to the {74,911} patches in \modisHAC{} to obtain a set of $k^{\ast}$ cluster centroids, $\mu=\{\mu_1, \dots, \mu_{k^{\ast}} \}$, where $k^{\ast}$ is the number of clusters defined in Section~\ref{sec:evaluation}.
We next want to assign a cluster label to each of the 198 million patches in \ocpatches{}.
We do this by identifying for each patch $x_i$
the cluster centroid $\mu_k$ with the smallest  Euclidean distance to its latent representation, $z(x_i)$.
We use Euclidean distance as our metric because our HAC algorithm uses Ward's method with Euclidean distance. 
That is, we calculate the cluster label assignment $c_{k,i}$ for the $i$-th patch as:   
\begin{equation}\label{eq:inference}
    c_{k,i} =  \argmin_{k=\{1,\dots, k^{\ast}\} }{|| z(x_i)  - \mu_k ||_2 } .
\end{equation}

This label prediction or \textit{{inference}} process is easily parallelized.
We use the Parsl parallel Python library~\cite{babuji2019parsl}, which enables scalable execution on many processors via simple Python decorators, for this purpose.
We observe an execution time of {533} seconds per day of MODIS imagery ($\sim${13,000} patches) on 256 cores of the Argonne Theta supercomputer.

\section{Evaluating Cluster Stability}\label{sec:stability}

Cluster stability is an important property for a cloud classification algorithm~\cite{kurihana2019cloud}.
A clustering method is said to be \textit{stable} for a dataset, $D$, and a number of clusters, $k$, if it produces similar or identical clusters when applied to different subsets of $D$.
As noted in Table~\ref{tab:fiveprotocols}, we define four tests to evaluate this criterion:

\begin{enumerate}
    \item     We measure \textit{{clustering similarity}} by generating clusterings for different subsets of the same dataset, and calculating the average distance between those clusterings. 
    \item     We measure \textit{{clustering similarity significance}} by comparing each clustering similarity score to that obtained when our clustering method is applied to data from a uniform random distribution. 
    \item    We measure \textit{{intra-cluster texture similarity}} by calculating the average distance between latent representations in each cluster. 
    \item     We measure \textit{{seasonal stability}} by comparing intra-cluster texture similarity for patches from January and July.    
\end{enumerate}

We are concerned not only to determine whether our clustering method, RICC, generates clusters that are stable, but also to identify the optimal number of clusters, $k^{\ast}$, to use for AICCA.
In determining that number, we must consider all four tests just listed: we want a high clustering similarity, a high significance (certainly greater than 1), a low intra-cluster similarity score, and low intra-seasonal texture differences. 

For all of our stability tests, we work with $D$ = \{\ocpatches{} from 2003 to 2021, inclusive\}. 
$|D| \approx 180$ M. 
(We do not consider data from 2000--2002 because 
Terra and Aqua were not operating at the same time for an entire year-long observation during that period.) 
We create a holdout subset $H$ with number of patches $N_H$ = {14,000}, and create 30 random subsets $S_i$ with $N_R$ = {56,000} by sampling without replacement from $D\setminus H$. This procedure ensures that the different $S_i$ are mutually exclusive and that there is no intersection between our holdout set $H$ and the random subsets. The ratio $N_H$ : $N_R$ of 20 : 80  is standard practice. 
We then create our 30 test datasets as $H \cup S_i$ for $\forall i \in \{1,\dots, 30\}$.

\begin{figure}[!htp]
    \centering
    \begin{minipage}{0.85\columnwidth}
        \subfloat[\label{fig:similarity_ari}]{\includegraphics[clip, width=.8\textwidth]{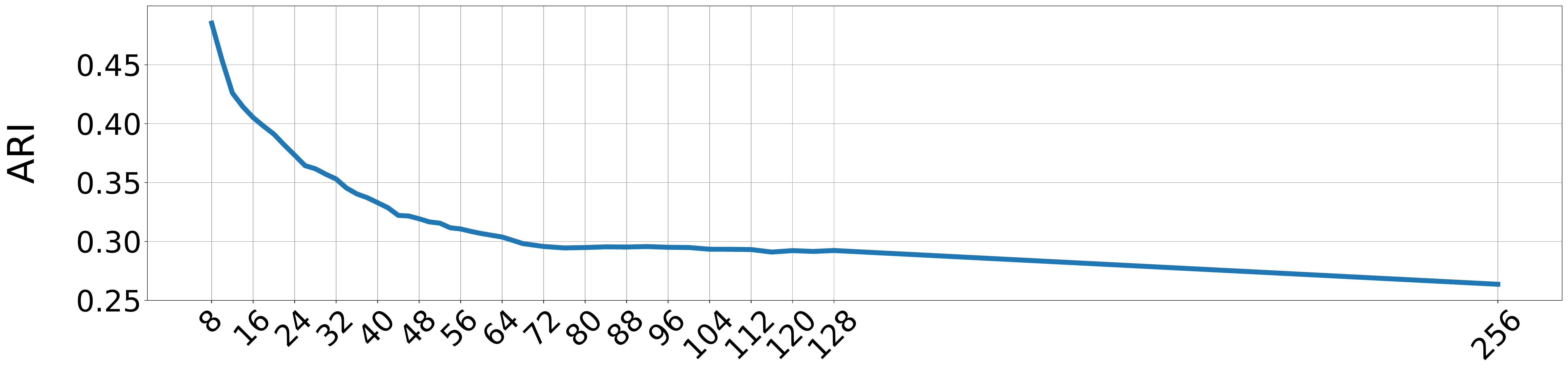}
        }
    \end{minipage} \\
    \begin{minipage}{0.85\columnwidth}
        \subfloat[\label{fig:significance}]{\includegraphics[clip, width=.8\textwidth]{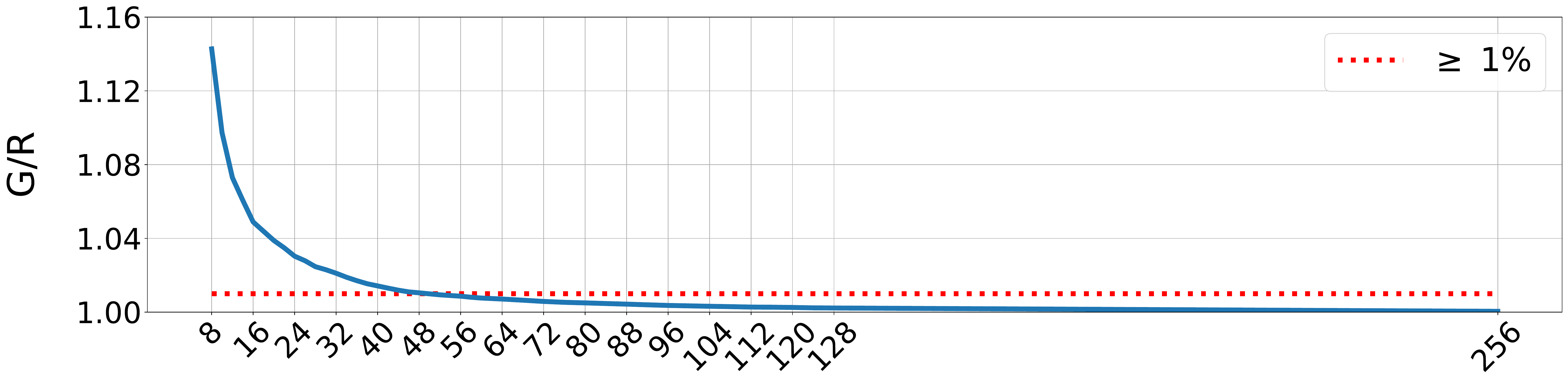}
        }
    \end{minipage}\\
    
%
%

    \begin{minipage}{0.85\columnwidth}
        \subfloat[\label{fig:pairwise_distance_weighted}]{\includegraphics[clip, width=.8\textwidth]{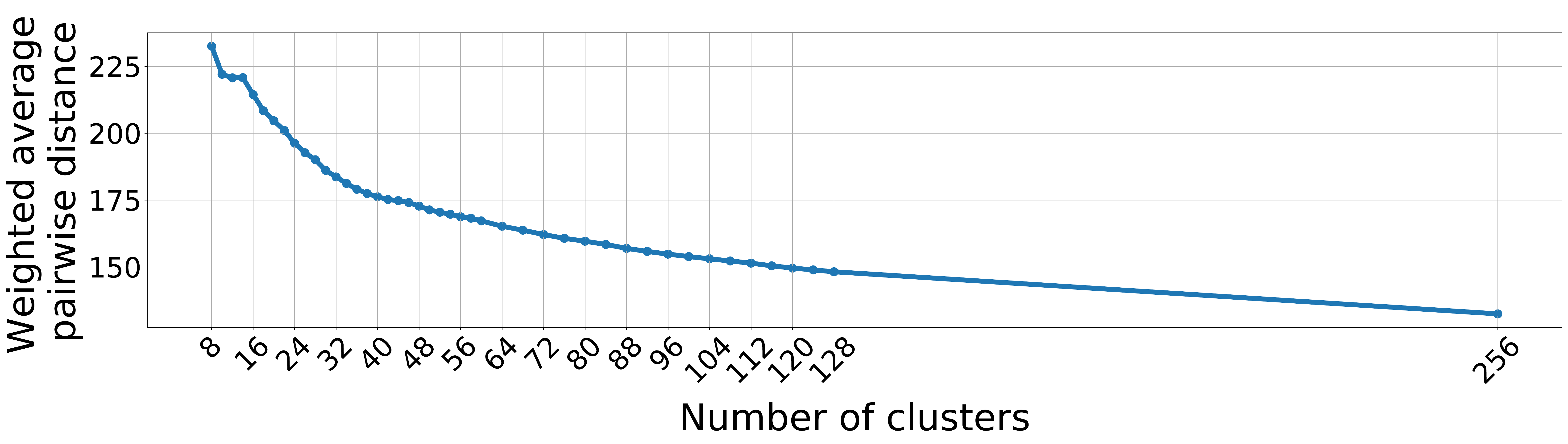}}      
    \end{minipage}
 \caption{Plots for the first three stability criteria metrics of Table~\ref{tab:fiveprotocols}, each as a function of the number of clusters. 
 (\textbf{a}) \textit{{Clustering similarity}}: 
 Adjusted Rand Index (ARI) as a measure of similarity of clusterings generated by RICC models trained on different subsets of patches. 
 (\textbf{b}) \textit{{Clustering similarity significance}}: 
 The blue line represents the ratio of the mean Rand Index based on RICC applied to our holdout patches $\{x \mid x \in H\}$ (G) to the mean Rand Index from HAC applied to random uniform distributions (R). The red dashed line is G/R $\geq$ 1.01, indicating that the stability of cluster label assignments produced from RICC is $\geq$1\% better than results of simply clustering random uniform data. 
 (\textbf{c}) \textit{{Intra-cluster texture similarity}}: 
 The blue line shows the weighted average of the mean squared Euclidean distance between pairs of patches within each cluster. Lower values suggest more homogeneous textures and physical features within each cluster. 
 The use of three similarity tests allows for achieving both stability and maximality criteria when grouping clusters.
 }\label{fig:stability_results}

\bigskip
    \centering
    \includegraphics[width=.8\columnwidth,trim=1mm 1mm 1mm 1mm,clip]{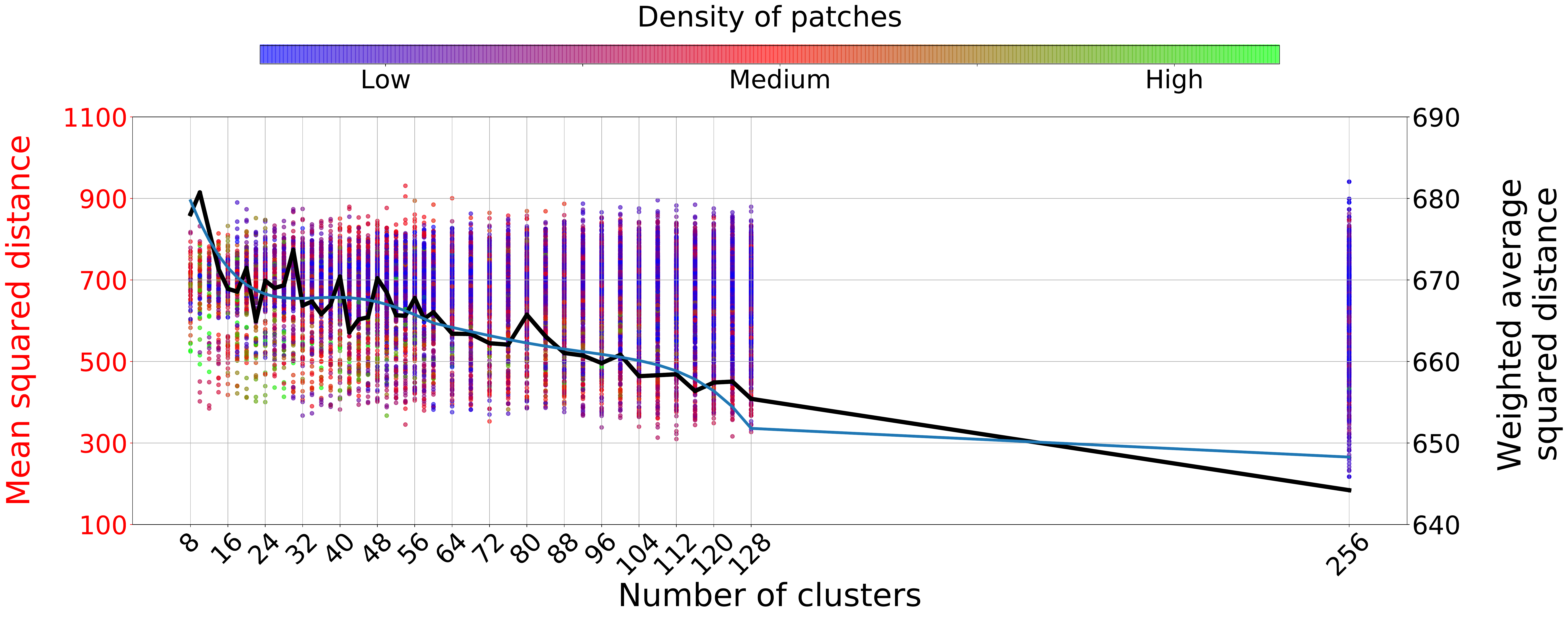}
    \caption{{Seasonal stability test comparing the intra-seasonal variance of textures within each cluster as a function of the number of clusters.}
    Each of $9\cdot k$ colored dots for each value of $k$ gives the average squared distance (left $y$-axis) between July and January patches as described in the text; 
    the color indicates cluster density, a measure of cluster size.
    The black line shows the mean WASD 
    (right $y$-axis) from nine trials as described in text. 
    The blue line shows a smoothed WASD curve obtained by applying a Savitzky--Golay filter with a degree six polynomial.
    The minimum WASD value in 
    $40 \leq k^\ast \leq 48$ occurs at $k$ = 42, motivating our choice for AICCA.
    }
    \label{fig:wasd}
\end{figure}

In the remainder of this section, we describe four stability tests, whose results are shown in Figures~\ref{fig:stability_results} and~\ref{fig:wasd}. These tests lead us to choose 42 as the optimal number of clusters. We also conduct additional evaluations of whether the result of using RICC with 42 clusters creates cloud classes that have reasonable texture and physical properties, when compared to similar exercises with suboptimal numbers of clusters.

\subsection{Stability Test 1: Clustering Similarity} \label{sec:stabilitytest1}

We measure clustering similarity by first generating clusterings for different subsets of the target dataset and then calculating the average pairwise distance between those clusterings. This approach is documented as Algorithm~\ref{alg:stability1} in Appendix~\ref{sec:apdx_clustering_similarity}.
As described above, we work with sets $H \cup S_i$, $i \in 1..30$, to generate 30 different clustering assignments via a trained RICC.
We compute the adjusted Rand index, ARI (Appendix~\ref{sec:similarity_scores}), as a measure of pairwise distance between pairs of clusterings. We average among the 30 clusterings generated by the models $\{ \text{RICC}_k^{i}$, $i \in 1..30 \}$ to determine the mean clustering similarity for that specific cluster number $k$, and then calculate the ARI for all $\binom{30}{2} = 435$ combinations of those 30 clusterings to determine the mean ARI score $G$. See Appendix~\ref{sec:apdx_clustering_similarity} for details.

The optimal number of clusters $k^{\ast}$ should have $G > 0$, and 
a higher score indicates that patches are more stably grouped into the same clusters. 
Figure~\ref{fig:stability_results}a shows that the mean ARI drops from 0.48 at eight clusters to 0.32 at 48 clusters, and then continues to decline to below 0.3 after 68 clusters.
Although the ARI score of 0.32 with 48 clusters is far from the perfect score of 1, previous literature~\cite{santos2009use} on the relative association of ARI scores and supervised learning measures for multiclass datasets reports that an ARI of 0.29 corresponds to 63.13\% in the classification correct percentage rate (COR) in the configuration of supervised learning, and that an ARI of 0.46 corresponds to 62.4\% in COR.
In addition, visual inspection suggests that the clusters produced by the RICC stably group similar cloud patterns.

\subsection{Stability Test 2: Significance of Similarities}\label{sec:stabilitytest2}

Having determined how cluster similarity scores vary with the number of clusters, we next turn to the question of whether these values are significant. 
Following Von Luxburg~\cite{von2010clustering}, we compare cluster similarity scores, as shown in Algorithm~\ref{alg:stability} {in Appendix~\ref{sec:apdx_stability_significance}}, against those obtained when the same method is applied to data generated not by our trained autoencoder but from a random uniform distribution clustered with the same HAC method. 
{We then compute the mean clustering similarity score $G$ from our patches and $R$ from the data from the random uniform distribution for each $k$ for all 435 combinations, though here we use the Rand index (as described in Appendix~\ref{sec:similarity_scores}) rather than ARI, as we are not comparing scores across $k$.}
We can then compare how the ratio between those two values varies with number of clusters.
A ratio $> 1$ indicates that cluster assignments are more stably grouped than would be expected by chance; a value of 1 indicates that there is no benefit to adding extra clusters.

We expect the ratio 
$G/R$ to be more than 1 if RICC cluster assignments are more stable than than those obtained on the null reference distribution.
We set a threshold of $G/R \geq 1.01$, meaning that the results obtained with RICC should be 1\% or more better than those with the null distribution.   
Figure~\ref{fig:stability_results}b shows the significance of the stability values $G/R$ as a function of the number of clusters $k$.
The significance curve drops to 1.01 at 50 clusters, indicating an optimal cluster number $k^\ast < 50$.

\subsection{Stability Test 3: Intra-Cluster Texture Similarity}\label{sec:stabilitytest3}

A stable clustering should group patches with similar textures within the same cluster.
To determine whether a clustering has this property,
we examine how the average distance between latent representations within each cluster changes when we apply RICC to create different numbers of clusters.
The mean distance between pairs of latent representations in a cluster relates to their similarity of texture, as our RI autoencoder learns texture features and encodes those features in latent representations.  
Specifically, we calculate the mean squared Euclidean distance between the latent representations computed for patches in our holdout set $H$.

For a clustering with $k$ clusters, let $n_c$ be the number of elements in cluster $c$, and $y_1${ ..} $y_{n_c}$ be the patches in that cluster.
As cluster sizes can vary,
we weight each cluster's mean distance by $w_c = n_c / \sum_{i=1}^{k} n_i$, to obtain a weighted average mean squared Euclidean~ distance:

\begin{equation}\label{eq:intradistance}
    d_{k} =  \sum_{c=1}^{k} \left({w_c}\sum_{i=1}^{m}\sum_{j>i}^{m} \frac{|| z(y_i)  - z(y_j) ||^2_2 }{\frac{m}{2}(m-1)}\right)\ \ \text{where}\  m = \text{min}(n_c, N_p), 
\end{equation}
where $z$ represents the latent representations generated by our RI autoencoder, and $N_p$ is the maximum number of patches to consider in the distance calculation---a limitation used to accelerate calculations. 
We set $N_p$ = 200 for our tests.
Note that, when the total number of clusters is large, some individual clusters may have a size less than this limit.


We calculate Equation~(\ref{eq:intradistance}) for $k$ from 8 to 256 for each of our 30 clusterings of test subsets $\{\text{RICC}_k^{1}(H),\dots, \text{RICC}_{k}^{30}(H)\}$, and then compute the mean value across clusterings. The resultant weighted average distance decreases monotonically with the cluster number $k$: see Figure~\ref{fig:wasd}c, as does the metric $G/R$ from test 2, but the trends have opposite implications: lower values are worse in test 2 but better in test 3.
A lower distance value indicates that cloud texture and physical properties are more homogeneous within a given cluster, meaning the resultant AICCA dataset provides a more consistent cloud diagnostic.
The implication is that the optimal number of clusters $k^{\ast}$ will be approximately the largest number that satisfies our criterion in test 2.

In Figure~\ref{fig:wasd}c, 
the distance metric sharply decreases from 8 to 36 clusters, but the slope then flattens and values are almost unchanged 
between 40--48 clusters. That is, the pairwise similarity of latent representations drastically increases between 8 and 36 clusters but becomes less different among the range between 40--48 clusters. Selection of a $k$ value from within this range would not change the result significantly. 
Since test 2 provides an upper bound of  $k^\ast < 50$, the results of test 3 suggest that 
the optimal number of clusters lies in $ 40 \leq k^\ast \leq 48$. 

To summarize: 
We observe that, as $G/R$ decreases, ARI also declines, and that our $G/R$ threshold requires $k^\ast < 50$.
We observe that a cluster number in the range $40 \leq k \leq 48$ satisfies all four stability criteria. 
We have validated that these choices also satisfy criteria 1--4 in Table~\ref{tab:fiveprotocols}.


\subsection{Stability Test 4: Seasonal Variation of Textures within Clusters}\label{sec:seasonal_clusters}
The results of the three tests above indicate that choices in the range $ 40 \leq k^\ast \leq 48$ will yield clusters that not only are stably assigned but also group similar cloud texture patterns.
Our final test investigates whether clusters produced via RICC show similar patterns regardless of season:
we compare intra-cluster texture similarity between \ocpatches{} from January and July. 
If differences are small, the number of clusters used is sufficient to accommodate the large seasonal changes in cloud morphology.


We use RICC with the autoencoder trained on \modisAE{} and cluster centroids based on \modisHAC{}, for different numbers of clusters $k$, as before.
For each $k$, we then apply the trained RICC$_k$ model to the patches in \modisHAC{} to assign a label $c \in \{1, .., k \}$ to each patch, and for each $c$, extract the latent representations for $m^s_c$ randomly selected July patches and $m^w_c$ randomly selected January patches with that label (with $m^s_c$ and $m^w_c$ being at most 100 in these analyses, but less if a particular cluster has fewer January or July patches, respectively), compute an intra-cluster texture similarity score for each set of July and January patches, and (as in Section~\ref{sec:stabilitytest3}) weight each cluster mean by the actual $m^s_c$ or $m^w_c$ so that we can consider texture similarities from many clusters without results being dominated by trivial clusters that we observe to group fewer similar patches due to undersampling.
We then sum the scores to obtain the overall weighted averaged squared distance (WASD) for $k$ clusters. In summary:

\begin{equation}\label{eq:wasd}
    \text{WASD}_k =  \sum_{c=1}^{k}\left( {w_c}\sum_{i=1}^{m^s_c} \sum_{j=1}^{m^w_c} \frac{|| z(y^s_i)  - z(y^w_j) ||^2_2 }{m^s_c \cdot m^w_c} \right) 
\end{equation}
where $w_c$ and $z$ are as defined in
Section~\ref{sec:stabilitytest3} and
$y^s$ = \{$y_1^s$ .. $y_{m^s_c}^s$\} 
and
$y^w$ = \{$y_1^w$ .. $y_{m^w_c}^w$\}
are the January and July patches in cluster $c$, respectively. 


We expand the analysis to account for two additional potential sources of bias. Because the specific days used in \modisHAC{} may affect our results, we assemble two additional versions of \modisHAC{}, selecting two days without replacement from each season in 2003, as before. The resulting \modisHACtwo{} and \modisHACthree{} have {77,235} and {76,143} patches, respectively.
Similarly, to account for any effect of the random selection of the $m^s$ summer and $m^w$ winter patches, we repeat the analysis of Equation~(\ref{eq:wasd}) three times for each of \modisHAC{}, \modisHACtwo{}, and \modisHACthree{}.
In this way, we obtain a total of $9\cdot k$ mean squared distance values and nine WASD values for each $k$ in the range 8 to 256. These are shown as the dots in Figure~\ref{fig:wasd}. The WASD curve (black) decreases with increasing cluster number $k$, implying as expected that higher cluster numbers allow for better capturing of seasonal changes. Because a smoothed version of the WASD curve (blue) has a minimum of $k$ = 42 over the range $40 \leq k \leq 48$, we choose 42 clusters as the optimum number and use this value in the inference step of Section~\ref{sec:inference}. 
Given that the WMO cloud classes define approximately 28 subcategories, the 42 AICCA clusters should not overwhelm users who use AICCA to investigate cloud transitions.


\subsection{Sanity Check: Comparison of RICCs with Different Number of Clusters to ISCCP Classes}\label{sec:isccp_modis_association}

As a final step, to confirm the utility of the choice of 42 classes, we consider whether and how RICC clusters associate with the nine ISCCP classes. We compare and contrast the frequencies of co-occurrence of (a) RICC clusters and (b) ISCCP classes, and evaluate how this relationship varies with cluster number used, considering not only the selected $k$ = 42 but also $k$ = 10, 64, and 256.

Recall that each of the nine ISCCP classes is defined by a distinct range of cloud optical thickness (COT) and cloud top pressure (CTP) values~\cite{rossow1999advances}: high, medium, and low clouds, and thin, medium, and thick clouds.
To compare RICC clusters with ISCCP classes, we calculate the relative frequency of occurrence (RFO) of RICC clusters across the same two-dimensional COT--CTP space, a standard approach to evaluating unsupervised learning algorithms~\cite{tselioudis2013global,mcdonald2018comparison,schuddeboom2018regional}.   
For this evaluation, we use the cluster assignments obtained with RICC when trained on \modisAE{} and \modisHAC{} to produce the AICCA dataset, as described in Section~\ref{sec:method}.
We take the Terra satellite ocean-cloud patches for January and July 2003,
and for each cluster, use the mean and standard deviation of the COT and CTP values for its patches to define a rectangular region for that cluster within two-dimensional COT-CTP space that extends for one standard deviation on either side of the mean.
We then calculate the number of clusters that are associated with each of the nine ISCCP classes by counting the number of clusters that overlap with that region of COT-CTP space and dividing this number by the total number of cluster-class overlaps for all clusters and classes.
Note that the latter number will typically be greater than the number of clusters because a single cluster can extend over multiple ISCPP classes.

This analysis shows a similar proportionality between RICC unsupervised learning clusters and ISCCP observation-based classes.
Table~\ref{tab:phys_rfos} compares the resulting proportions of RICC clusters (for each value of cluster number $k$) with the simple mapping of all patches to ISCCP classes based on their COT and CTP values (top line).
In all cases, the Stratocumulus (Sc) class is the largest single category, and medium-thickness clouds (Sc, As, Cs) predominate at each altitude level.

Stratocumulus (Sc) account for approximately 30\% of RICC cluster overlaps, while the proportion of cloud observations in this category is over 50\%.  Similarly, for all $k$ values, relatively few RICC clusters are assigned to high clouds, as expected since these make up only \textasciitilde15\% of total cloud occurrences. The thin and medium ISCCP classes (Cu, Sc, Ac, As), which account for 78.4\% of cloud occurrence in the MODIS dataset, are represented by a similar proportion of RICC cluster overlaps: 74.45\%, 70.44\%, and 71.40\% for $k$ = 42, 64, and 256 clusters, respectively. 
There is no physical reason that cluster overlaps and cloud occurrence frequencies need be exactly the same: if, for example, all low medium-thickness clouds were identical in texture, we would expect that they would be assigned to a single cluster. However, the similarity of proportions suggests that AICCA captures physically meaningful distinctions among cloud types.

\subsection{Discussion of Stability Protocol Results}\label{sec:discussStability}

We have used the stability protocol described in this section to determine the number of clusters that both achieves a stable grouping of patches and maximizes the richness of the information contained in our clusters.  
Recall that Von Luxburg's normalized stability protocol~\cite{von2010clustering} simply minimizes an instability metric to determine the number of clusters that maximize stability.
In contrast, 
we combine four tests---adjusted cluster similarity, normalized stability, weighted intra-cluster distance, and seasonal texture differences---to address the stability criterion.
We used these tests to evaluate whether the cloud clusters produced by our unsupervised learning approach can provide meaningful insights for climate science applications.

This use of multiple similarity tests is essential to achieving our goal of both stability and maximality when grouping clusters.
The \textit{{clustering similarity}} test gives a mean score of scaled values calculated by ARI as a measurement of the degree of stability in \ocpatches.    
While this value is easy to understand when the resulting mean ARI is close to 1 (i.e., \ocpatches{} are always clustered into the same cluster group), ARI when applied to real world data could result in a value that is close to neither 0 nor 1~\cite{santos2009use}.

\begin{table}[t!]
\centering
\caption{
ISCCP: Relative frequencies of occurrence based on mean COT and CTP values for \ocpatches{} from January and July, 2003. 
AICCA: Relative frequencies of occurrence of RICC clusters over each of the nine ISCCP cloud classes~\cite{rossow1999advances}, as determined by counting the number of clusters that overlap (as determined by the mean, plus or minus one standard deviation, of COT and CTP values for patches within each cluster) with each class, divided by the total number of cluster-class overlaps. We allow double counts if a cluster overlaps more than one ISCCP class.  
Results are given for $k$=10, 42, 64, and 256 clusters, and for just January patches, just July patches, and both January and July patches.
The AICCA values that are closest to the frequencies from MODIS column are in boldface.
Recall that MODIS values are based on frequencies of \textit{patches} over COT-CTP space, while the AICCA values are based on frequencies of \textit{clusters} over COT-CTP space.
Note that frequencies in each line add to 100, modulo rounding.
We observe that the AICCA cluster frequencies are roughly proportional to the ISCCP category frequencies, 
although they consistently underestimate the Sc class (by 20\%), and overestimate Cu and As classes. 
}\label{tab:phys_rfos}
\begin{small}
\begin{tabular}{C{1cm}|C{0.8cm}|C{0.5cm}|r*{2}{r} | r*{2}{r} | r*{2}{r}  }
\hline
\multicolumn{3}{c|}{Height} & \multicolumn{3}{c|}{Low} & \multicolumn{3}{c|}{Medium} & \multicolumn{3}{c}{High} \\ \hline
 \multicolumn{3}{c|}{Thickness} & Thin & Med & Thick &
 Thin & Med & Thick &
 Thin & Med & Thick \\ \hline
 Dataset    & Month & $k$ & \multicolumn{1}{c}{Cu}  & \multicolumn{1}{c}{Sc} & \multicolumn{1}{c|}{St} & \multicolumn{1}{c}{Ac} & \multicolumn{1}{c}{As} & \multicolumn{1}{c|}{Ns} & \multicolumn{1}{c}{Ci} & \multicolumn{1}{c}{Cs} & \multicolumn{1}{c}{Dc} \\ \hline\hline
 \multirow{3}{1.1cm}{ISCCP}& \multirow{3}{1cm}{Jan \& July 2003} & & 
 & & & & & & & & \\ 
 & & & 5.29 & 53.94 & 2.93 & 3.65 & 15.50 & 2.12 & 3.39 & 10.52 & 2.60  \\
 & & & & & & & & & & &  \\\hline\hline
\multirow{12}{1cm}{AICCA$_k$} 
    & \multirow{4}{1cm}{Jan 2003}
      & 10	&  11.42	& 25.71	& 8.57 & \bftab 5.71 & \bftab 22.85 &	8.57 &	2.85 &	8.57 &	5.71 \\
    & & 42	& 12.50	& 29.16 & 4.16 &  10.00  & 25.00 & 5.00 &	2.50 &	8.33 &	3.33 \\
    & & 64	& 10.38	& \bftab 34.41 & \bftab3.24 & 7.79 & 26.62 & \bftab 3.24 & 0.64 & \bftab 9.74 & 3.89 \\
    & & 256	& \bftab 9.06 & 32.90 & 4.45 & 8.90 & 25.27 &  5.08 &  \bftab 3.65 &	 8.58 & \bftab 2.06 \\ \cline{2-12} 
    & \multirow{4}{1cm}{July 2003}
      & 10	&  13.33 & 23.33 & 3.33 & \bftab 6.66  & \bftab 16.66 &	3.33 & 6.66 & 16.66 & 10.00 \\
    & & 42	& 10.30	& 30.92 & 4.12 & 7.21  & 20.61 & 4.12 & \bftab 3.09 & \bftab 13.40 &	6.18 \\
    & & 64	& 10.20	& 30.61 & 1.36 & 10.20  & 19.04 & 2.72 & 5.44 & 14.28 &	6.12 \\
    & & 256	& \bftab 9.31 & \bftab 32.16 &\bftab 2.46 & 8.78 & 21.61 & \bftab 1.93 & 5.97 & 13.53 & \bftab 4.21 \\ \cline{2-12} 
    & \multirow{4}{1cm}{Jan \& July 2003}
      & 10	& 12.50	& 25.00	& 6.25 & 3.12 & \bftab18.75 & 3.12 & 9.37 &	15.62 &	6.25 \\
    & & 42	& 10.67	& \bftab32.03 & \bftab 2.91 & 7.76  & 24.27 & \bftab 2.91 &  \bftab 2.91 &	\bftab 11.65 &	4.85 \\
    & & 64	& 8.80	& 29.55 & 3.77 & 8.17  & 23.89 & 5.03 & 5.03 & 11.94 & \bftab 3.77 \\
    & & 256	& \bftab8.42	& 31.57 & 3.63 & 7.93 & 23.47 & 3.96 & 5.45 & 11.73 & 3.80 \\ \hline 
\end{tabular}
\end{small}
\end{table}

The \textit{{significance of similarities}} test enables us to find the number of clusters after which there is reduced merit, from the perspective of stability against the null reference distribution, in adding more clusters. 
Normalized stability thus provides statistical support for eliminating certain cluster numbers, especially when the first test produces an ARI value that is close to neither 0 nor 1.

We introduce the \textit{{similarity of intra-cluster textures}} test because common approaches to estimating an optimal number of clusters, such as the elbow method~\cite{bholowalia2014ebk}, silhouette method~\cite{rousseeuw1987silhouettes}, and gap statistics~\cite{tibshirani2001estimating}, seek to determine the \textit{minimum} number of clusters needed to characterize a dataset, 
which is not our goal. In our application, achieving a minimum number of clusters might result in the merging of sub-clusters
with 
unique textures and slightly different physical properties.
By minimizing the intra-cluster difference shown in Figure~\ref{fig:wasd}c until the slope of the curve of distance becomes small,
the third test causes the lower bound on the optimal number of clusters to increase to $40 \leq k^{\ast}$, avoiding oversimplifications.

Finally, the \emph{{seasonal stability}} test provides a further validation of our choice of $k^*$. A too-small number of clusters is likely to result in dissimilar July or January patches being mapped to the same cluster. We see in Figure~\ref{fig:wasd} a local minimum in weighted average intracluster seasonal difference.

A disadvantage of our stability protocol is that, unlike other heuristic approaches~\cite{tibshirani2001estimating,von2010clustering}, it does not always determine a unique optimal number of clusters.
Indeed, our stability protocol in Section~\ref{sec:stability} concludes that $40 \leq k^{\ast} \leq 48$. 
Determining a single optimal number in the range sandwiched by the results of the four tests ultimately requires a subjective choice, based for example on the structure of cloud clusters in \modisHAC.
In this study, we chose 42 as the number in the range $40 \leq k^{\ast} \leq 48$ that minimizes the seasonal variation of textures within clusters: see Section~\ref{sec:seasonal_clusters}---although we note that a different selection of \ocpatches{} in \modisHAC{} could motivate a different value. 

\section{Results}\label{sec:results}
Having determined in Section~\ref{sec:stability} an optimal number of clusters, $k^{\ast}$, 
we then validate the scientific utility of AICCA\textsubscript{42} by evaluating the relationship between cloud class labels and their physical properties and spatial patterns. 
We have previously verified that the cloud clusters produced by RICC are physically reasonable using a limited subset of the MODIS data~\cite{kurihanaRICC21}.
This section provides a similar analysis on a far more complete dataset of {589500} Terra ocean-cloud patches for January 2003 and July 2003. The goal is to confirm that AICCA\textsubscript{42} diagnoses meaningful physical properties for use in climate science applications.

\subsection{Seasonal Variability of Cloud Cluster Regimes}\label{sec:regime} 

\begin{figure}[h]
    \centering
    \begin{adjustwidth}{-0cm}{0cm}
        \centering
       \includegraphics[clip, width=1\textwidth]{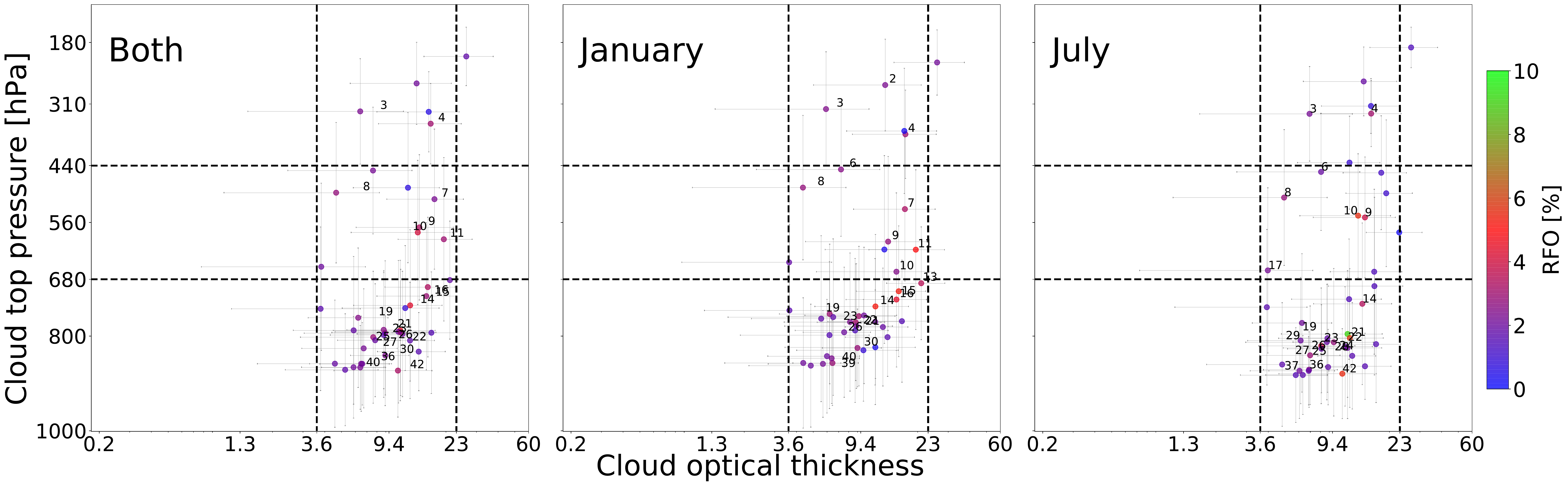}
 \end{adjustwidth}
 \caption{Distributions of cluster properties for AICCA\textsubscript{42} in
 {COT--CTP} space, where COT is cloud optical thickness (dimensionless) and CTP cloud top pressure (hPa). We show January and July 2003 (left), January only (center), and July only (right). Dots indicate mean values for each cluster and error bars the standard deviation of cluster properties. Data point colors indicate the relative frequency of occurrence (RFO) of each individual cluster in the dataset. 
 Note that, in assigning cluster labels, we sort the clusters first on CTP and then on the global occurrence of the clusters within each 50 hPa pressure bin.
 Thus, small cluster numbers (e.g., \#1) represent high-altitude cloud, and within a similar CTP range (e.g., 500 hPa--550 hPa), smaller numbers represent the more dominant patterns within the bin. 
 For clarity, we show only the 21 clusters with the highest RFOs. For comparison, dashed lines divide the COT-CTP space into the nine regions corresponding to ISCCP cloud classes. 
 AICCA\textsubscript{42} captures a greater variety of cloud types than do the ISCCP categories, with most of the clusters at low altitude (high CTP). 
 January and July panels are similar, indicating that AICCA\textsubscript{42} adequately captures seasonal variation in cloud properties. 
 }\label{fig:physicalregimes}
\end{figure}

Because the Earth is not symmetric, its clouds show strong seasonal variability not only in any given location but in the global mean. In this section, we show that the physical properties of AICCA\textsubscript{42} clusters are reasonable and remain stable even if the dataset is restricted to a single month. 
This analysis builds on those in Sections~\ref{sec:seasonal_clusters} and~\ref{sec:isccp_modis_association}. In Section~\ref{sec:seasonal_clusters}, we used intra-cluster seasonal differences as a criterion for choosing an optimal $k$ of 42.
In Section~\ref{sec:isccp_modis_association}, we showed that RICC distributed those clusters in the COT-CTP space that defines established ISCCP classifications roughly in accordance with actual frequencies of cloud occurrence.  We now plot the cluster distribution in COT-CTP space, and show that it is indeed reasonably constant across seasons (Figure~\ref{fig:physicalregimes}).
Note that, in assigning cluster labels, we sort the clusters first on CTP and then
on the global occurrence of the clusters within each 50 hPa pressure bin.


As expected based on prior results, Figure~\ref{fig:physicalregimes} shows that most AICCA\textsubscript{42} clusters fall in the low cloud range (680--1100 hPa cloud top pressure) with low to medium optical thickness (2--20): Compare to Table~\ref{tab:phys_rfos}.
These results are broadly consistent with those of \citet{jin2017simplified}, who performed a simple clustering analysis with the joint histogram of optical thickness and cloud top pressure, though they obtained relatively more clusters associated with high clouds (four of their 11 clusters, vs.\ five of 42 in this work). The distribution of clusters is largely unchanged even when only January or July data are used in clustering. For example, the cumulus (Cu: left bottom) and stratocumulus (Sc: center bottom) regimes comprise 30 clusters in the full-year analysis, 30 in July only, and 32 in January only.

Using 42 clusters clearly allows RICC to capture richer cloud information than in the limited set of nine ISCCP cloud classes. In our previous work~\cite{kurihanaRICC21}, we found that 12 clusters were insufficient to achieve a clear separation between high and low clouds. In this work, the clusters from our cloud fields can distinguish the full range of physical properties here (from high to low CTP and thick to thin COT), though thin clouds are included only because our cloud clusters defined by means and error bars (i.e., standard deviation of the cloud parameter) cover more than one ISCCP class. 
The choice of a cluster number of 42 produces a reasonable trade-off.

\subsection{Comparing AICCA\textsubscript{42} and ISCCP Classifications}\label{sec:isccp_association}

\begin{figure}[!ht]
    \centering
    \begin{adjustwidth}{-0cm}{0cm}
        \includegraphics[clip, width=1\textwidth,trim=1mm 1mm 1mm 1mm]{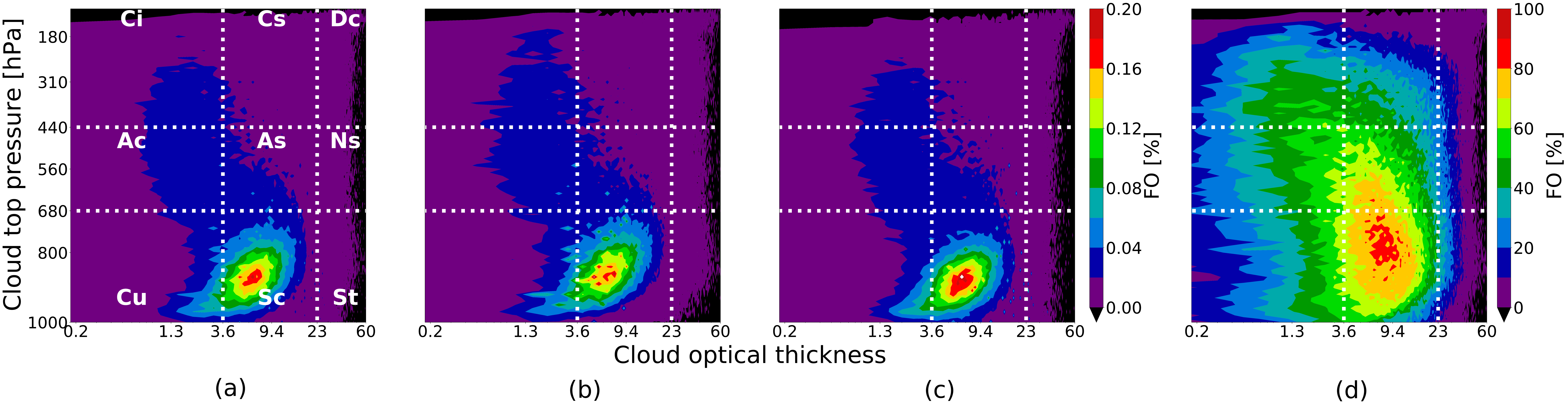}
    \end{adjustwidth}
 \caption{Heatmaps of the relative frequency of occurrence in COT-CTP space for (\textbf{a}) observed patches from both January and July, (\textbf{b}) values only from January, (\textbf{c}) values only from July, and (\textbf{d}) cluster counts inclusive of overlap from AICCA\textsubscript{42}.
 Distributions are smoothed; resolution is 0.5 for COT and 10 hPa for CTP. Panels for observed frequencies (\textbf{a}--\textbf{c}) and cluster density (\textbf{d}) are expected to have different values. For example, in (\textbf{a}), a heatmap value of 0.1\% indicates {5895} patches fall in a given histogram bin. In (\textbf{d}), a heatmap value of 71\% indicates that 30 of 42 clusters overlap with that histogram bin over the range of one standard deviation. 
 The data used here are those used throughout Section~\ref{sec:results}:  
 all ocean-cloud patches from January and July 2003 from the Terra instrument.   
 White dashed lines show the boundaries of the nine ISCCP cloud classes~\cite{rossow1999advances,cloudtypes}:
 Cirrus (Ci), Cirrostratus (Cs), Deep convection (Dc), Altocumulus (Ac), Altostratus (As), Nimbostratus (Ns), Cumulus (Cu), Stratocumulus (Sc), and Stratus (St).
 AICCA\textsubscript{42} clusters cover all nine ISCCP classes, with the largest representation in the Stratocumulus (Sc) category where occurrence also peaks.
 }\label{fig:aicca_isscp_hist2d}
\end{figure}

We now investigate further how AICCA\textsubscript{42} distributes clusters in COT-CTP space, and compare to observed occurrence frequencies. A limitation of the ISCCP cloud classification scheme is that the stratocumulus clouds whose behavior is of the greatest concern to climate scientists, and which comprise 54\% of the MODIS dataset (Table~\ref{tab:phys_rfos}), are lumped into a single ISCCP class (Figure~\ref{fig:aicca_isscp_hist2d}a--c). A major motivation for AICCA\textsubscript{42} is to provide greater interpretive detail for understanding these low, marine clouds. 



As shown in previous sections, AICCA\textsubscript{42} does provide a richer sampling of the stratocumulus (Sc) regime. AICCA\textsubscript{42} allocates 71\% of cluster centers to the stratocumulus regime (Figure~\ref{fig:physicalregimes}; 30 of 42 classes), or 32\% of their relative occurrence frequency inclusive of overlaps (Table~\ref{tab:phys_rfos}; see Section~\ref{sec:isccp_modis_association} for description of methodology). While Table~\ref{tab:phys_rfos} provided only mean values for each ISCCP class, Figure~\ref{fig:aicca_isscp_hist2d}d shows the full distributions. As we would hope,  AICCA\textsubscript{42} partitions cloud information more finely at low cloud altitudes and moderate cloud thickness (Sc), 
while still sampling every part of COT-CPT space.

\subsection{Separation of Ice and Liquid Phases}\label{sec:iceliquid}
\begin{figure}[!ht]
    \centering
    \begin{adjustwidth}{-0cm}{0cm}
    \includegraphics[clip, width=1\textwidth]{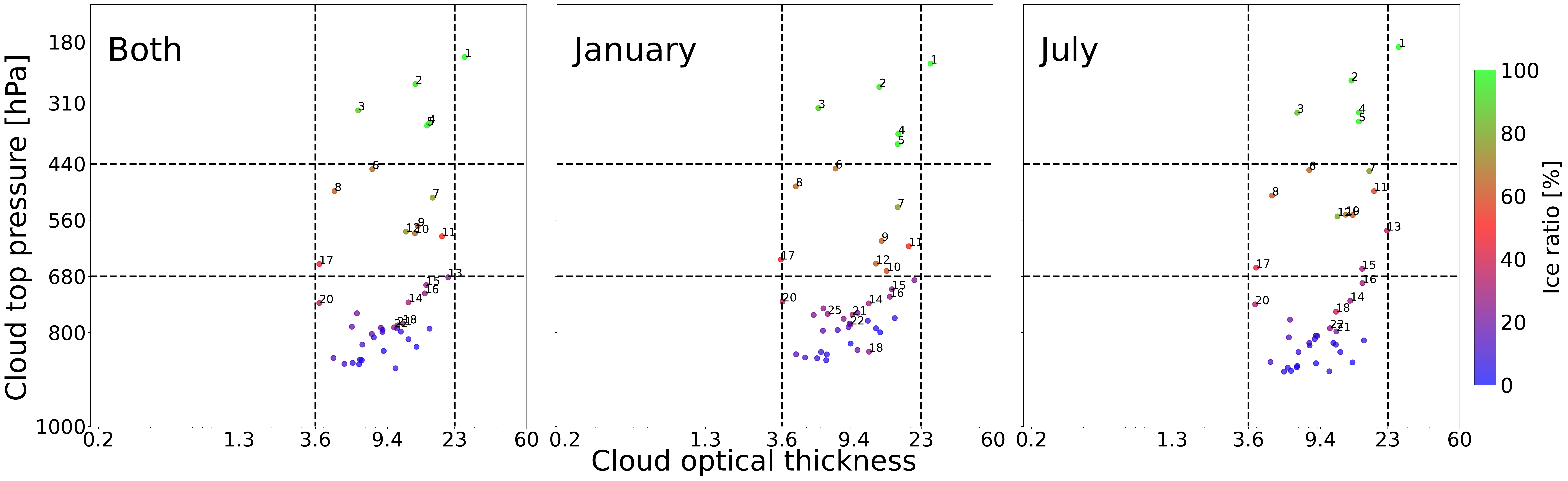}\label{fig:icean10}
\end{adjustwidth}
 \caption{Test of whether AICCA\textsubscript{42} captures expected variations in cloud microphysics, i.e., the ice and liquid fractions for individual cloud classes.
 The figure is constructed in the same way as Figure~\ref{fig:physicalregimes}, but with each color marker now showing the cluster's mean \textit{ice phase ratio}, defined as the mean within-cluster percentage of cloud pixels denoted as ice phase. We omit all pixels labeled as ``undetermined'' in MOD06; many of these are internally mixed phase but the proportions cannot be determined.
AICCA\textsubscript{42} cloud classes are sufficiently restricted that they capture the expected microphysics, with higher ice fractions in higher-altitude clouds.
 }\label{fig:iceregimes}
\end{figure}

We showed in previous work that RICC-generated clusters can differentiate between clouds that are dominated by ice vs. liquid phase. (See {Figure} 10 in Kurihana et al.~\cite{kurihanaRICC21}.) 
We extend this analysis here and demonstrate that the same discrimination occurs in the larger AICCA\textsubscript{42} dataset.  
Figure~\ref{fig:iceregimes} shows for each cloud class the average percentage of cloud pixels that are identified as an ice phase in the MOD06 cloud properties. As expected, cloud classes centered at high altitude (low CTP) are predominantly ice, those at middle altitudes are mixed, and those at low altitude are predominantly liquid. The lowest classes have $<$3\% ice labels, and note that MOD06 cloud properties themselves have some error rate.
The gradient in ice content across mid-level clouds, the region of transition from liquid to ice, also matches physical expectations. 
Note that while our ice phase ratio metric predominantly captures mixed-phase clouds, in which ice and liquid coexist in a single meteorological event (for our purposes, a patch), it is also affected by cases where a cluster contains a mix of pure-ice and pure-liquid clouds.

In summary, the AICCA\textsubscript{42} classes are sufficiently homogeneous to provide meaningful interpretation.
These results support the physical reasonableness of the AICCA dataset.


\subsection{Case Studies: Spatial Distribution of Cloud Textures and Associated Cluster Labels}\label{sec:clusterexample}

\begin{figure}[!hp] 
    \vspace{-2ex}
    \centering
    \begin{adjustwidth}{-0cm}{0cm}
    \begin{minipage}{.35\columnwidth}
        \subfloat[\label{fig:geoswath}]{\includegraphics[clip, width=0.99\textwidth]{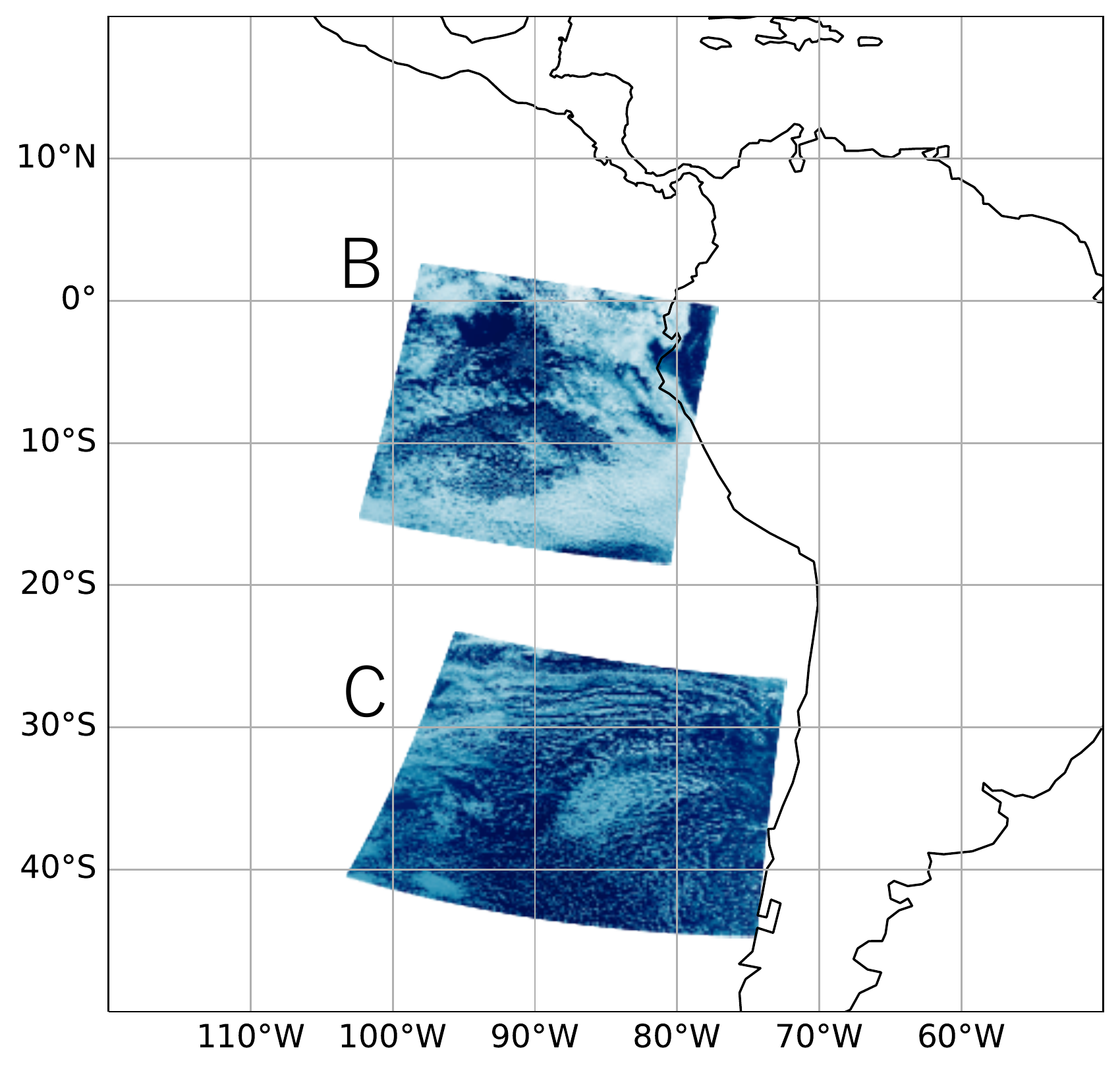}}
    \end{minipage} 
    \begin{minipage}{.28\columnwidth}
        \subfloat[\label{fig:aicca42swath1}]{\includegraphics[clip, width=0.99\textwidth]{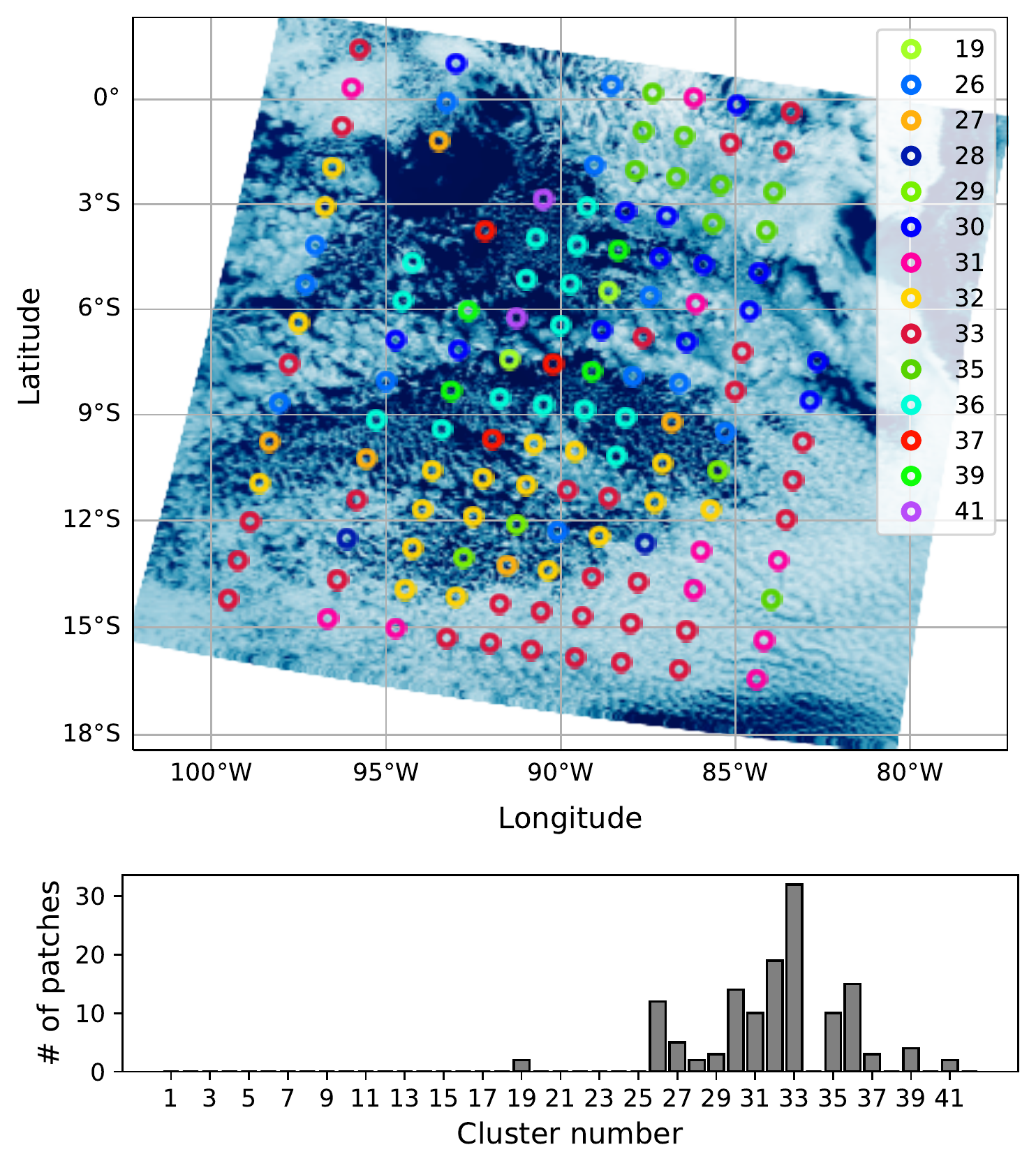}
        }
    \end{minipage} 
    \begin{minipage}{.28\columnwidth}
        \subfloat[\label{fig:aicca42swath2}]{\includegraphics[clip, width=1.17\textwidth]{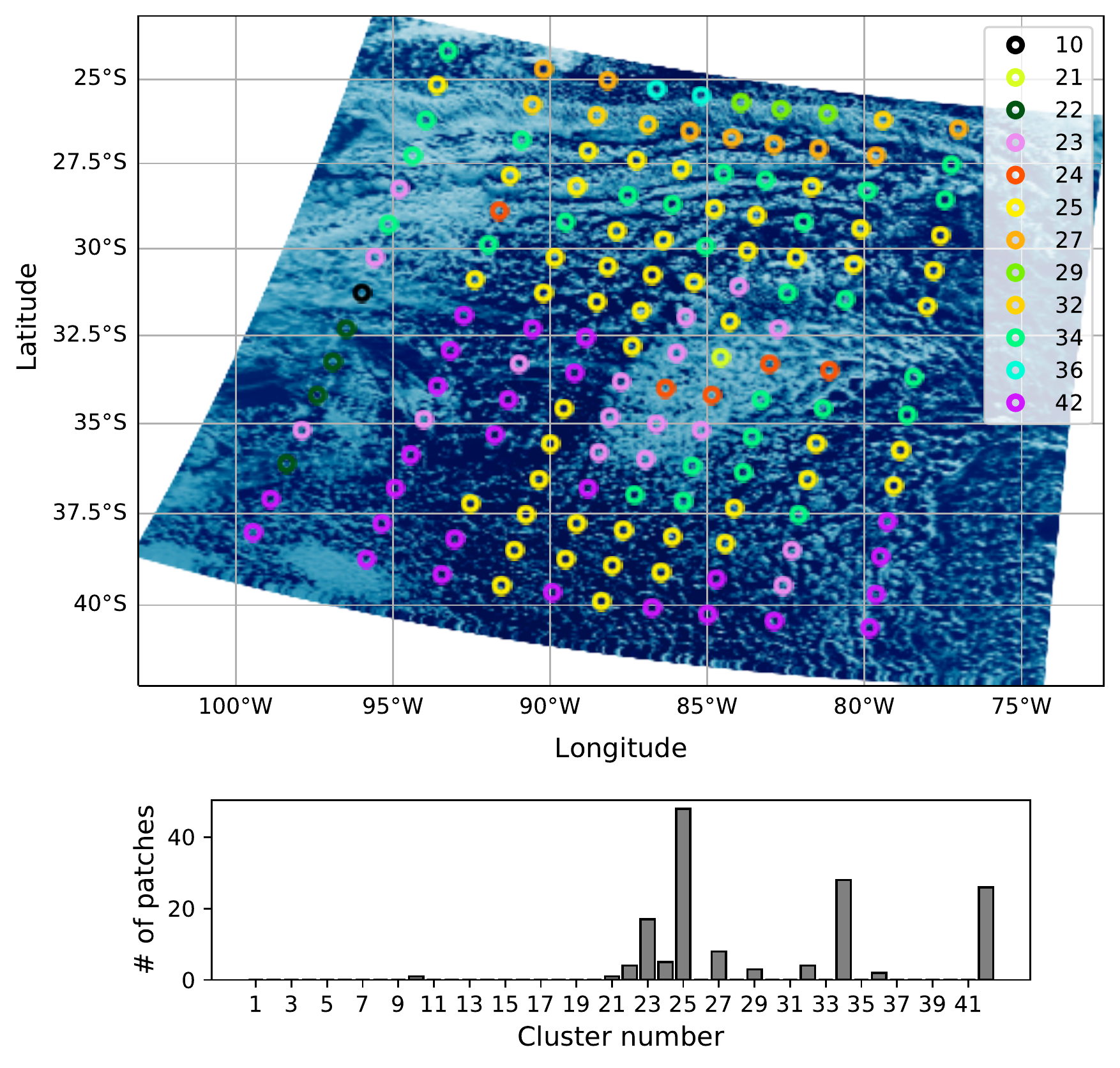}
        }
    \end{minipage} \\
    \vspace{-1ex}
    \begin{center}
        \begin{minipage}{1\columnwidth}
        \subfloat[\label{fig:swathtexture}]{\includegraphics[clip, width=\textwidth]{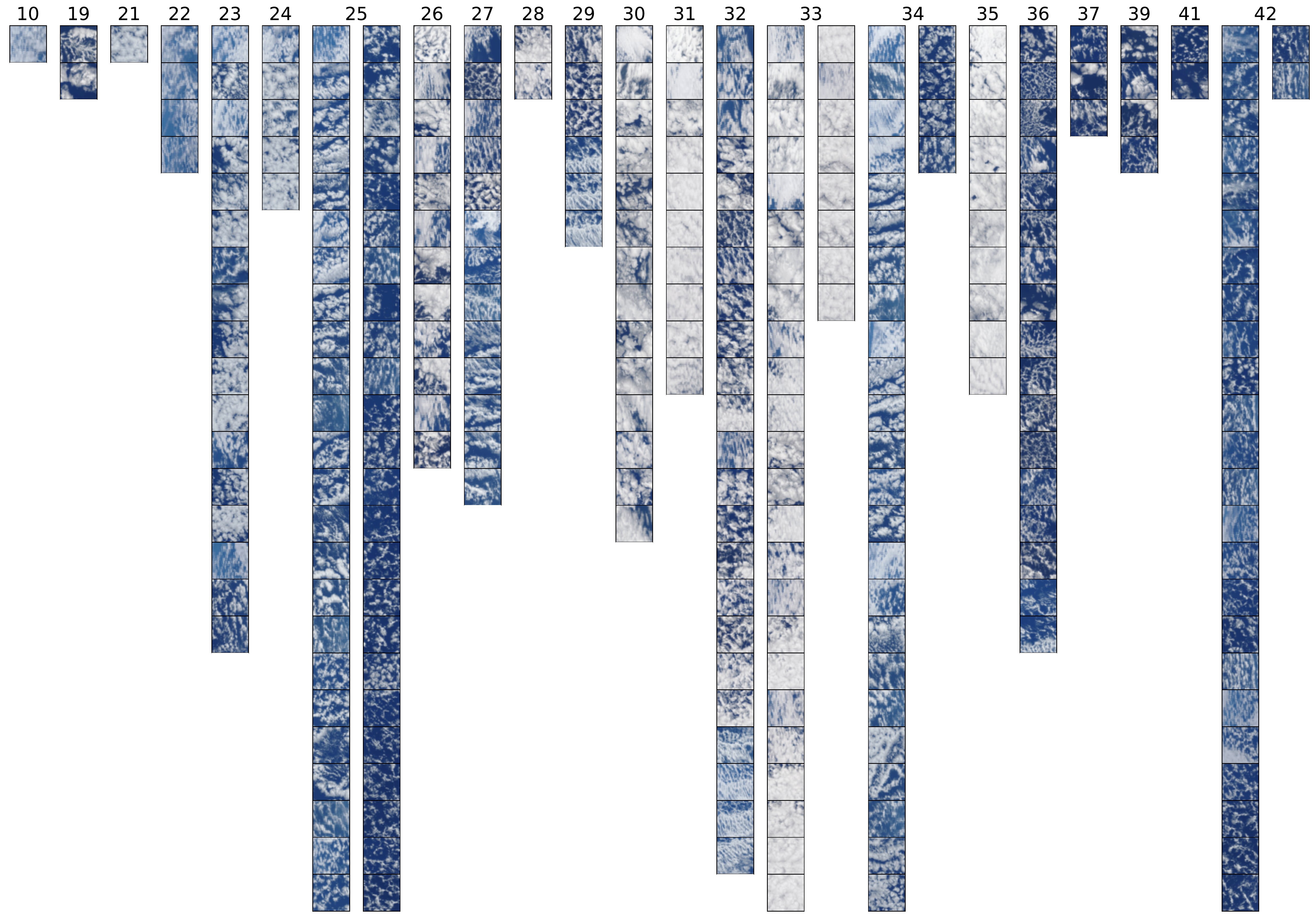}}
        \end{minipage}
    \end{center}
    \end{adjustwidth}
    \vspace{-1ex}
    \caption{
    (\textbf{a}) Geographical location of two example MODIS swaths (B and C) off the west coast of South America, both from the Terra instrument but at different times;
    (\textbf{b}) Swath B, {133} ocean-cloud patches between 18$^\circ$ S to 3$^\circ$ N, 76$^\circ$ S to 104$^\circ$ S, observed on 15 January 2003, with each patch represented by a dot with color indicating its associated class label in the range 1..42; 
    (\textbf{c}) Swath C, {147} ocean-cloud patches between 44$^\circ$ S to 23$^\circ$ S, 72$^\circ$ S to 103$^\circ$ S, observed on 20 July 2003, similarly labeled. Note that not all clusters appear in each swath. Histograms in (\textbf{b}) and (\textbf{c}) show the distribution of cloud class labels; note there is little overlap;
    (\textbf{d}) MODIS true color images~\cite{gumley2003creating} for all ocean-cloud patches labeled in (\textbf{b}) and (\textbf{c}), grouped by cluster number. Note the visual similarity of cloud textures within each cluster.
    AICCA\textsubscript{42} produces spatially coherent cluster assignments, groups visually similar textures, provides rich detail by subdividing stratocumulus clouds into multiple classes, and identifies subtle spatial and/or temporal differences.    
    }
    \label{fig:example_swath}
\end{figure}

To provide a visual example of the power of AICCA\textsubscript{42} classes in interpreting cloud processes, we examine two case studies involving swaths of MODIS imagery, both dominated by marine stratocumulus,
off the west coast of South America: see Figure~\ref{fig:example_swath}a. Note that the swath labeled B is from January and that labeled C from July. 

The two example swaths show the richness and diversity of stratocumulus patterns. The more equatorial swath B (Figure~\ref{fig:example_swath}b) shows regions of both open- and closed-cell stratocumulus clouds, and sharp transition regions. 
The mid-latitude summertime swath C (Figure~\ref{fig:example_swath}c) is dominated by open-cell stratocumulus clouds, with broad transitional regions and only small patches of classic closed-cell.

AICCA\textsubscript{42} cluster labels capture important aspects of these distributions.
As usual, we label only patches with $>$30$\%$ cloud pixels; each such patch is marked with a dot in Figure~\ref{fig:example_swath}b,c, with the color denoting the cluster label. 
The cloud classes assigned are geographically contiguous and reflect clear visual distinctions in cloud texture (Figure~\ref{fig:example_swath}d). 
They also capture important and subtle distinctions. Each swath contains 12--14 unique classes, but only four are shared between both. That is, cloud classes of otherwise similar visual appearance are strongly differentiated in space and/or time. Open-cell stratocumulus in swath B is assigned to classes \#32 and \#36, but that in swath C largely to \#25, \#34, and \#42. Similarly, closed-cell stratocumulus in swath B is assigned to classes \#30, \#31, \#33, and \#35, none of which are present in swath C. Instead, the smaller areas of closed-cell stratocumulus in swath C are labeled as class \#24.
These results suggest that real-world stratocumulus cloud textures involve subtle but important spatial and/or temporal distinctions and that AICCA\textsubscript{42} is capturing those distinctions. 


\subsection{Use Case: Geographic Distribution of Cluster Label Occurrence}\label{sec:clusterexample2}

\begin{figure}[t] 
    \centering
    \begin{adjustwidth}{-0cm}{0cm}
        \includegraphics[width=18.5cm,trim=1mm 1mm 1mm 1mm,clip]{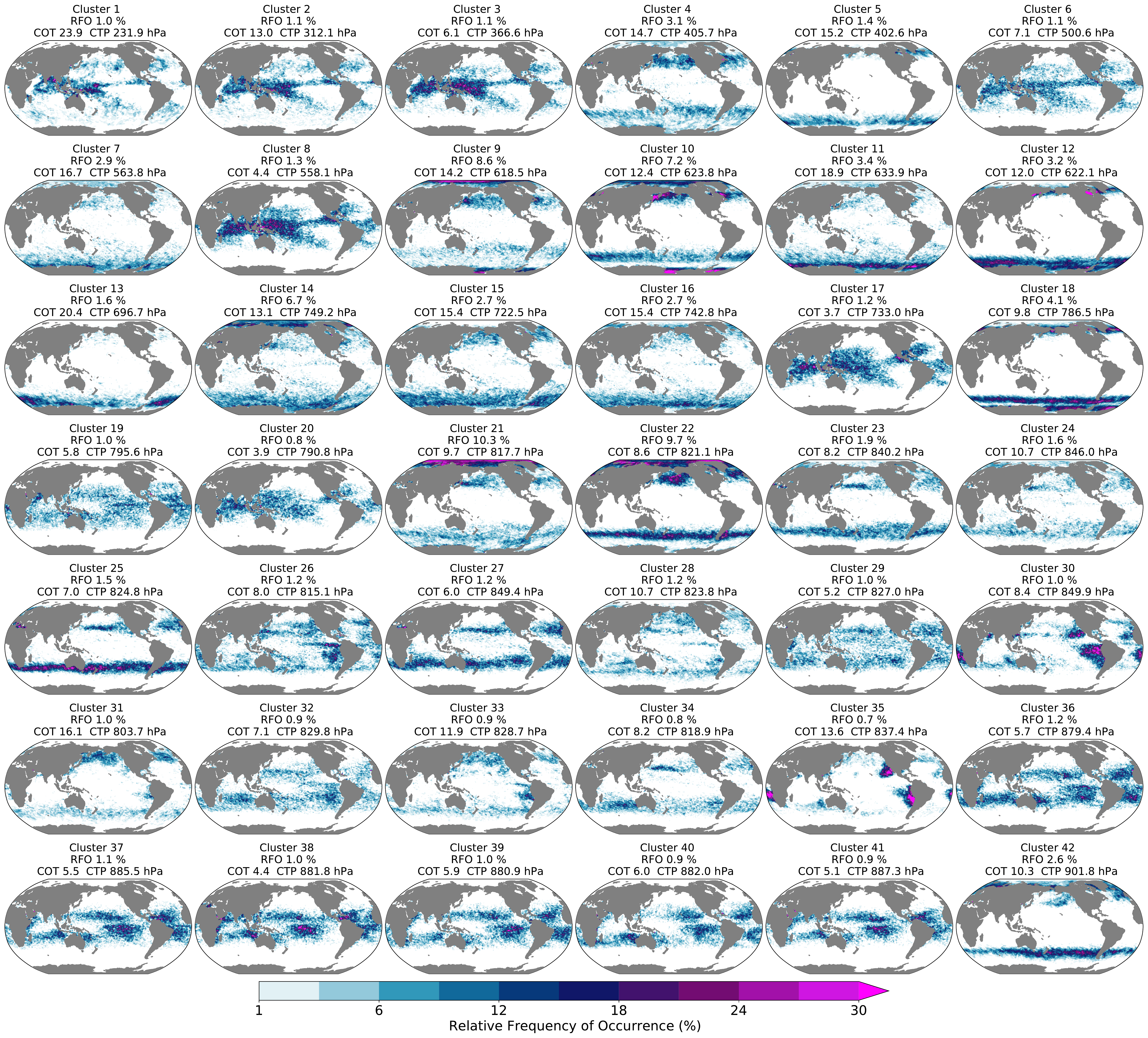}
    \end{adjustwidth}
    \caption{An example application of AICCA. We plot the relative frequency of occurrence (RFO) for each of the 42 AICCA\textsubscript{42} clusters, using all data from January and July, 2003. Land is in grey, and areas where RFO < 1.0\% are in white. 
    Surtitles show global mean RFO, cloud optical thickness (COT), and cloud top pressure (CTP) for the given cluster. 
    Clusters show striking geographic distinctions, and those with roughly similar spatial patterns have different mean physical properties, suggesting meaningful physical distinctions.
    The 99 percentile of RFO values (RFO $\geq$ 1 \%) of \#30 is 29.85 \% and the value of \#35 is 36.58 \%.
    }\label{fig:global}
\end{figure}

In this last study, we examine the geographic distribution of AICCA\textsubscript{42} cluster labels. Using the same dataset as in the other part of this section, 
we show in Figure~\ref{fig:global} mean incidences for each of the 42 cloud types in the dataset used throughout \autoref{sec:results}, gridded on a 1$^\circ$ global grid
We see strong geographic distinctions among cluster labels, with some occurring only in the tropics and others only at high latitudes. Some show even finer geographic restrictions. For example, cloud classes \#1--\#3 are localized primarily in the West Pacific warm pool, all likely associated with tropical deep convection, though ranging in altitude (232--367 hPa CTP) and thickness (24--6 COT). (Classes are numbered in order of their mean altitude; see Section~\ref{sec:regime} for details.)
By contrast, the stratocumulus cloud labels discussed for Figure~\ref{fig:example_swath} show different distributions. Those most clearly associated with classic closed-cell stratocumulus---\#30, \#33, and \#35---are as expected primarily localized to small areas on the west coasts of continents. The most predominant open-cell Sc cloud labels in Figure~\ref{fig:example_swath}---\#25, \#32, and \#36---are more widely distributed but with strong latitudinal dependence. 
The six clusters just described are all low in altitude (mean CTP of 803--901 hPa) and moderate in thickness (mean COT of 8.4--13.6 thickness for the closed-cell classes and 5.7--7.1 for the open-cell). All would therefore be labeled as Sc in the ISCCP classification; AICCA\textsubscript{42} reveals their striking differences. Note that, because our example dataset includes both January and July 2003, these graphs include both summer and wintertime occurrences. When displayed as an animation of monthly means, the geographic distinctions become even sharper, with patterns migrating seasonally with the sun's position.

\unskip

\begin{figure}[!h]
    \centering
    \begin{adjustwidth}{-0cm}{0cm}
    \begin{minipage}{.32\columnwidth}
        \subfloat{\includegraphics[clip,trim=2mm 2mm 2mm 2mm, width=0.90\textwidth]{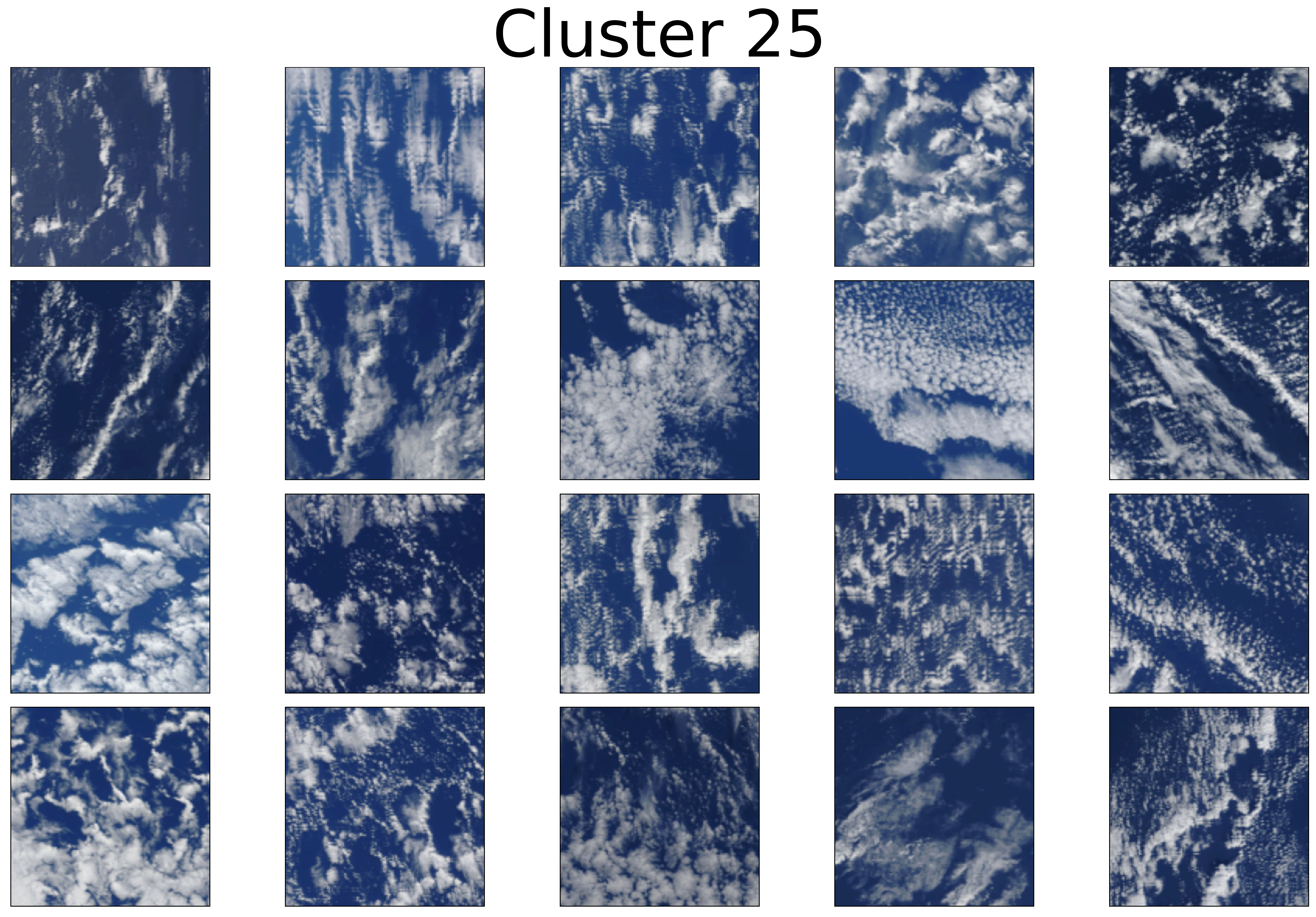}
        } 
    \end{minipage}
    \begin{minipage}{.32\columnwidth}
        \subfloat{\includegraphics[clip,trim=2mm 2mm 2mm 2mm, width=0.90\textwidth]{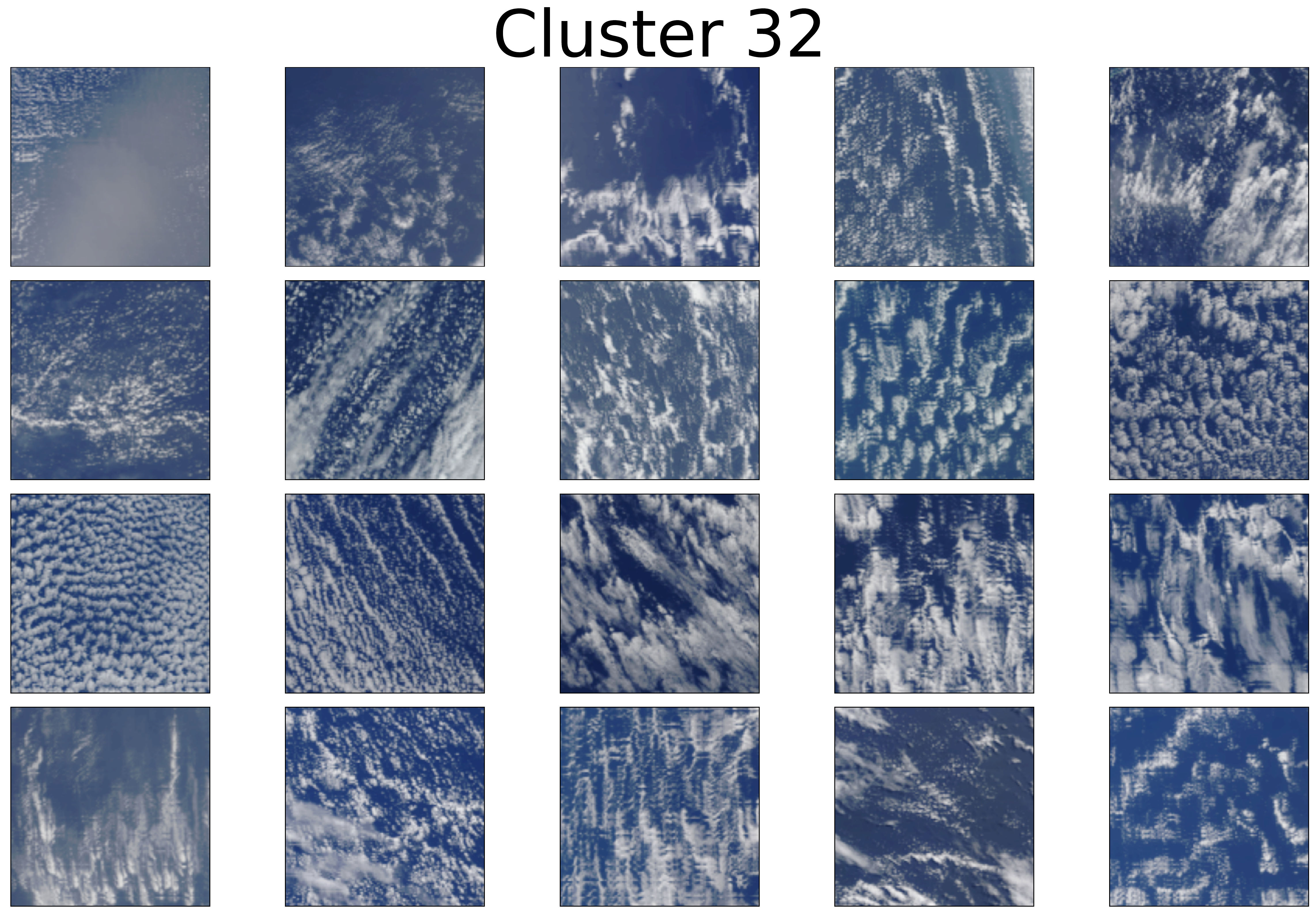}
        }
    \end{minipage}
    \begin{minipage}{.32\columnwidth}
        \subfloat{\includegraphics[trim=2mm 2mm 2mm 2mm,clip, width=0.90\textwidth]{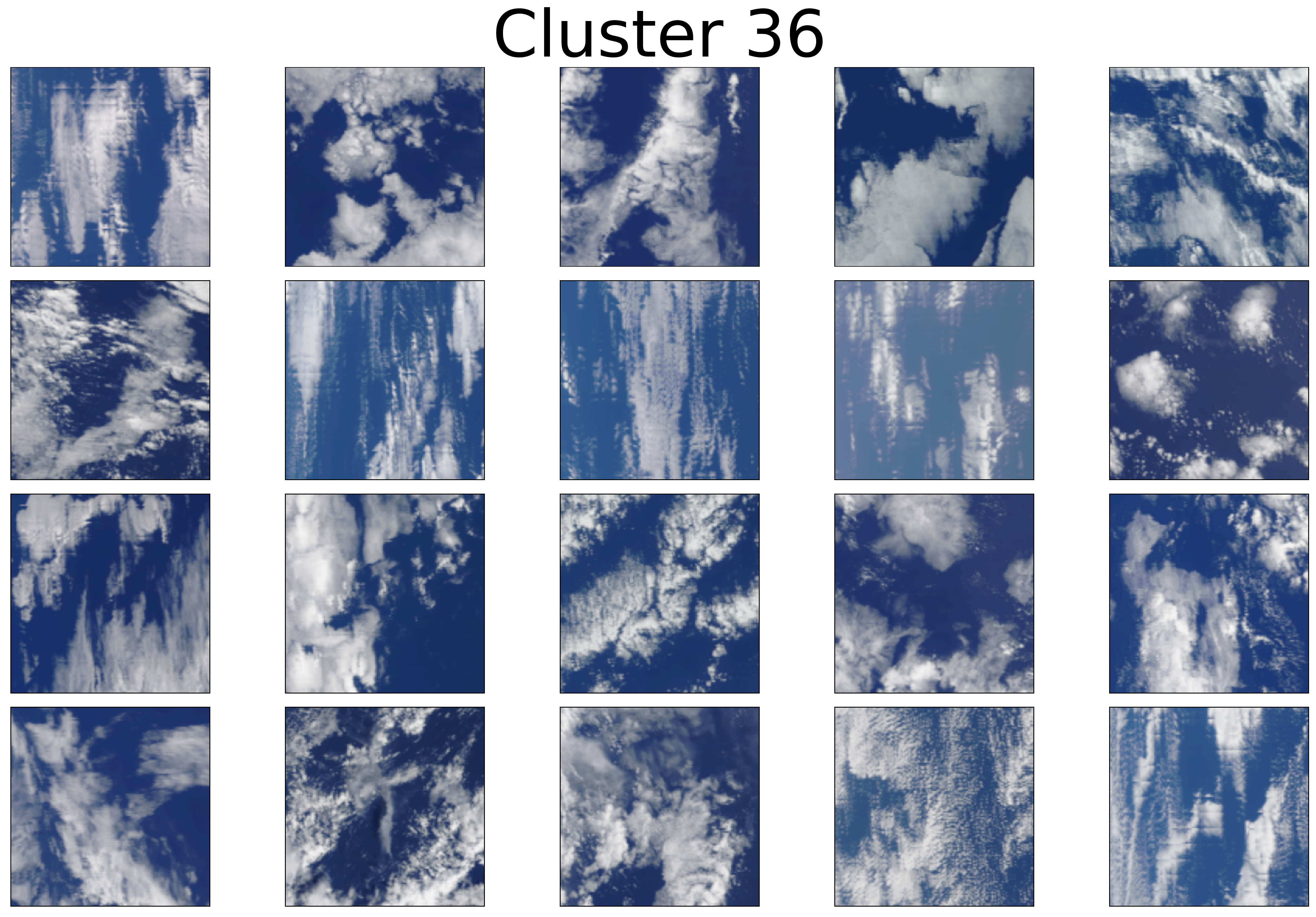}}
    \end{minipage}\\
    \begin{minipage}{0.32\columnwidth}
        \subfloat{\includegraphics[clip,trim=2mm 2mm 2mm 2mm, width=0.9\textwidth]{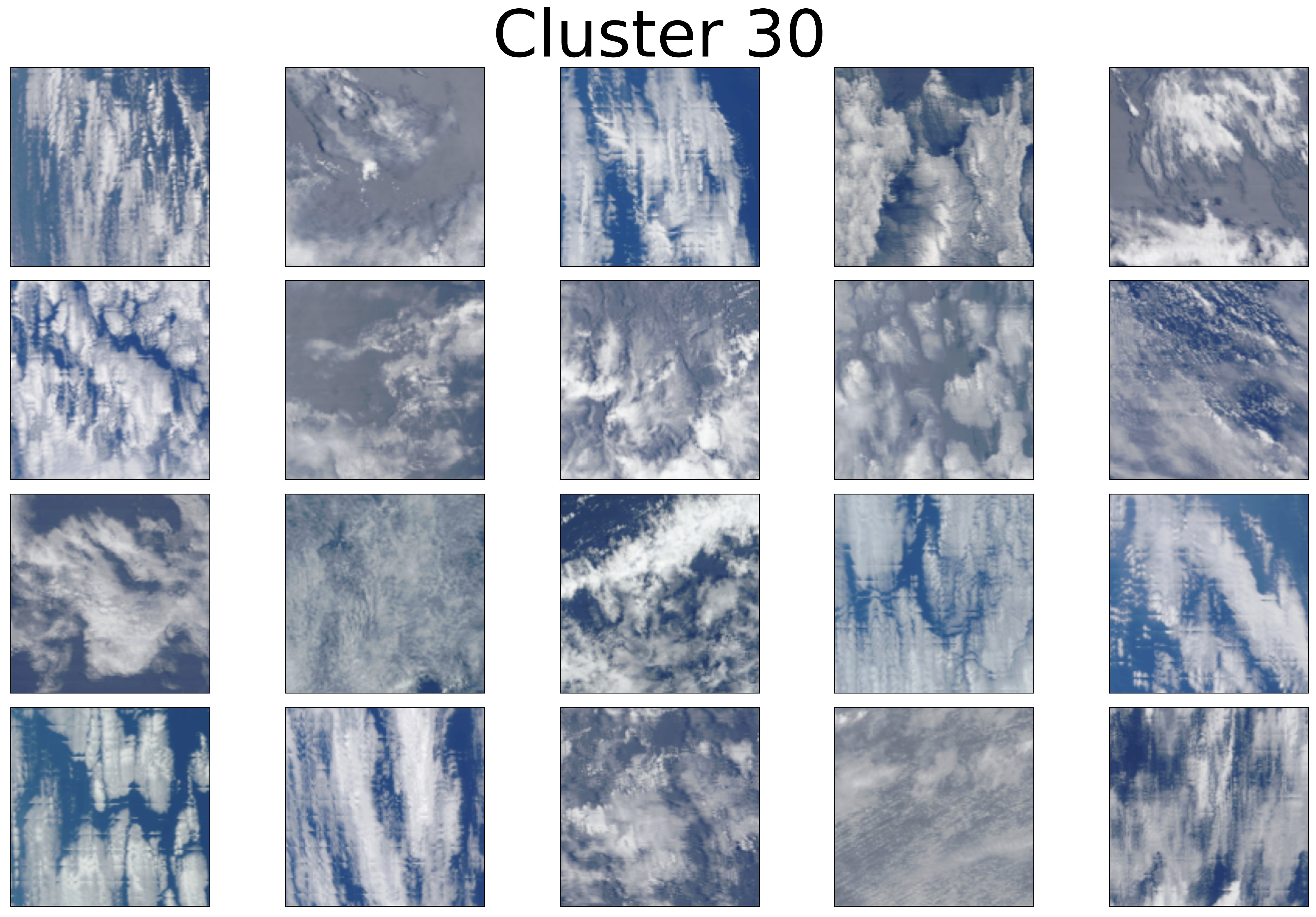}}
    \end{minipage}
    \begin{minipage}{0.32\columnwidth}
        \subfloat{\includegraphics[clip,trim=2mm 2mm 2mm 2mm, width=0.9\textwidth]{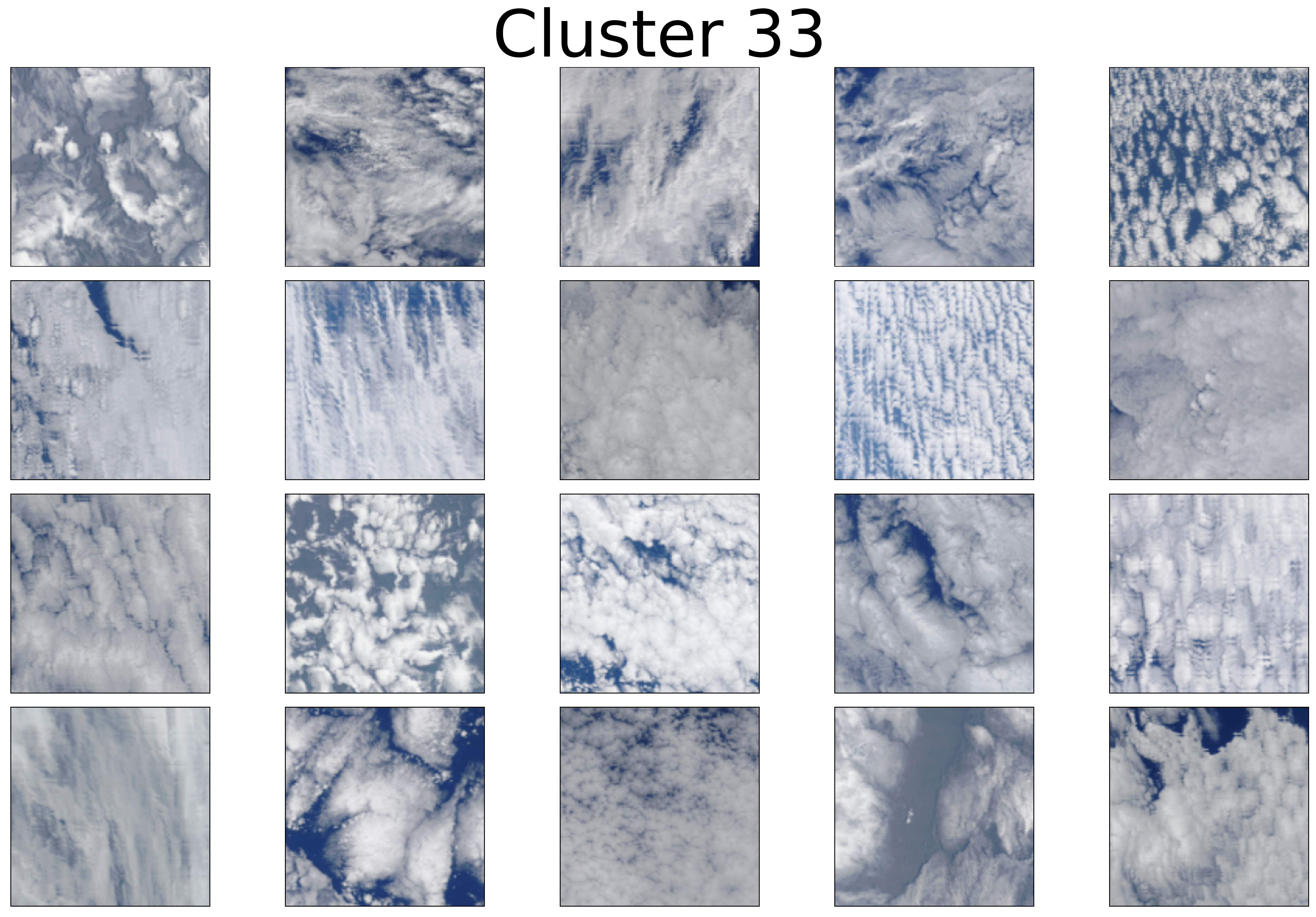}}
    \end{minipage}
    \begin{minipage}{0.32\columnwidth}
        \subfloat{\includegraphics[clip,trim=2mm 2mm 2mm 2mm, width=0.9\textwidth]{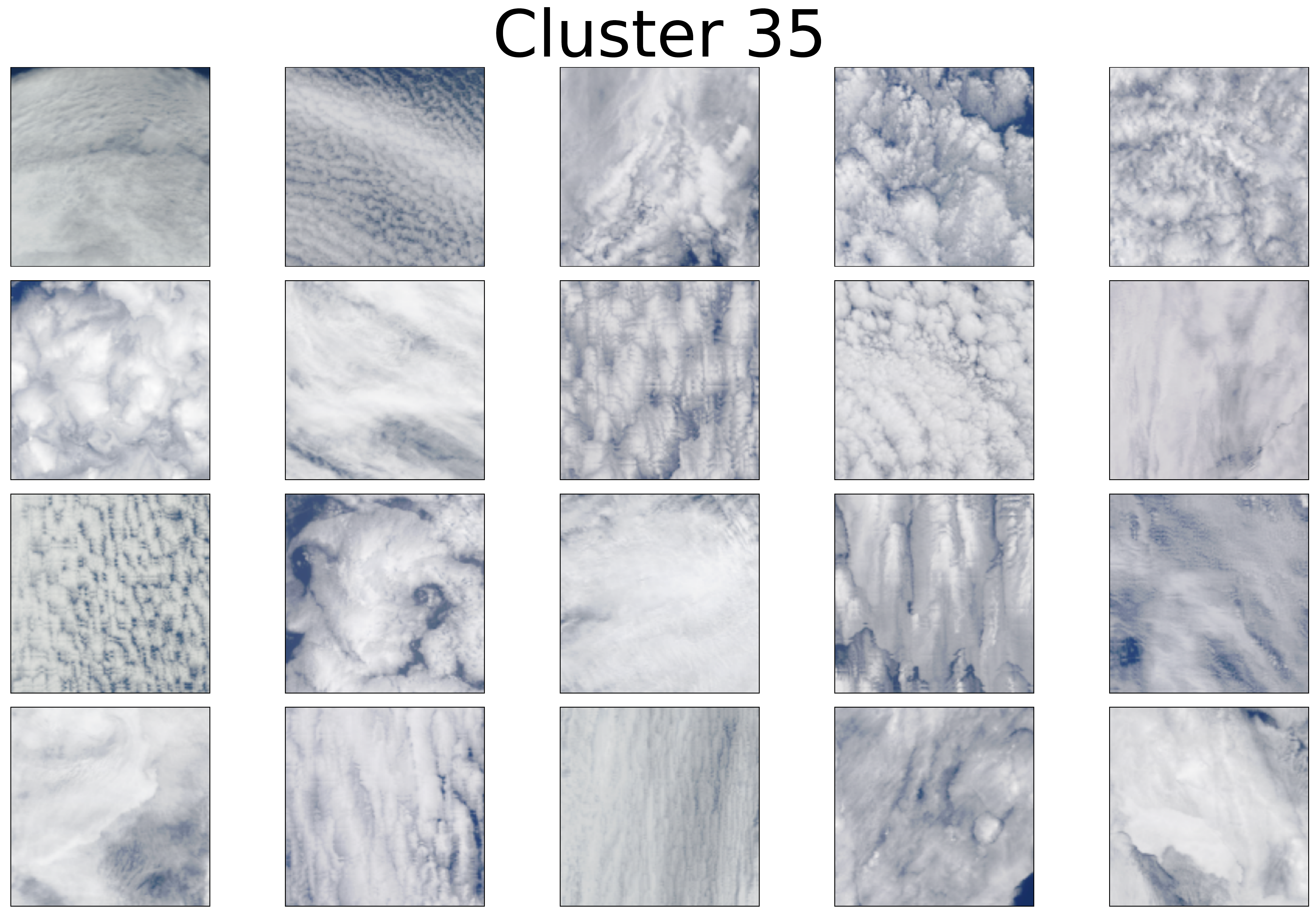}}
    \end{minipage}
    \end{adjustwidth}
    \caption{Selected MODIS true color images~\cite{gumley2003creating} for the six clusters that dominate open-cell (upper row: \#25, \#32, and \#36) and closed-cell (lower row: \#30, \#33, and \#35) stratocumulus clouds in Figure~\ref{fig:global}. Surtitles show the cluster numbers.
    We show the 20 patches closest to \modisHAC{} centroids. 
    Note how AICCA discriminates between textures (e.g., compare the fine-scale detail of \#32 to the more coarsely aggregated \#36) even for patches of similar mean cloud properties. 
   }
    \label{fig:top20}
\end{figure}

To highlight the {\it texture} distinctions in the Sc cloud classes just discussed, we show in Figure~\ref{fig:top20} the true color images~\cite{gumley2003creating} corresponding to the 20 patches closest to the \modisHAC{} centroid for each of the six clusters.
Patches shown for each cluster are visually similar, and the different clusters have distinct differences in not only cloud pixel density but also spatial arrangement, even within the broad open cell (top row, \#25, \#32, and \#36) and closed cell (\#30, \#33, and \#35) categories. These distinctions show that AICCA\textsubscript{42} is separating stratocumulus clouds by texture as well as by mean properties across the patch. 

The strong localization of some cloud classes near the poles raises concern that they may be affected by the presence of sea ice. We have restricted analysis to ocean clouds to avoid the complications of surface effects---the ocean provides a dark and homogeneous background---but parts of the high-latitudes ocean are covered in wintertime ice. Because two of the MODIS bands used in our cloud clustering system, bands 6 (1.6 $\upmu$m) and 7 (2.12 $\upmu$m), are also used by the MODIS snow and ice detection algorithm~\cite{riggs2015modis}, the resulting AICCA dataset can inadvertently include some surface information in the latent representation. 
To check for contamination, we use a MODIS cloud product that describes the presence of a snow and ice background for each pixel (MOD06). Only one cloud class may experience significant interference: \#12, which forms 
in local winter. (Sea ice makes up 16/31\% of its labeled pixels in January/July.) The other polar cloud classes 
appear in local summer. Sea ice effects therefore do not appear to drive the labeling of geographically distinct cloud classes that appear in polar oceans.

These results suggest that AICCA\textsubscript{42} identifies real and important differences between cloud types and can help climate scientists understand the drivers of distinct cloud patterns and regimes.


\section{Conclusions}\label{sec:conclusion}
We have introduced an AI-driven cloud classification atlas, AICCA\textsubscript{42} that provides the first global-scale unsupervised classification of clouds in MODIS satellite imagery. 
AICCA\textsubscript{42} provides a compact form of the information available in multi-spectral satellite images, reducing 801 TB of MODIS products to 54.2 gigabytes of cloud labels and, for diagnostic purposes, four cloud properties from MOD06 (cloud optical thickness, cloud top pressure, cloud phase, and cloud effective radius).
We have described the complete workflow used to generate the dataset, the five criteria used to assess its success (physically reasonable, spatial distributions, separable, rationally invariant, and stable), and the novel protocol developed to determine the optimal number of clusters that meets the stability~requirement.

The new stability protocol is needed because our goal differs from the norm in clustering studies, which generally seek to determine the {\it minimum} number of clusters needed to characterize a dataset. Instead, we seek to maximize the richness of information captured by determining the {\it maximum} number of clusters that remain stable to changes in the training set. The protocol of four tests suggests an optimal cluster number of $k^{\ast}$ = 42, and our seasonal stability sanity check confirms that this number is sufficient to capture the full seasonal diversity of global cloud textures. The resulting atlas of cloud classes greatly enhances the richness of information provided over the traditional 9-class ISCCP scheme, especially for climate-critical cloud types: for example, 30 of the AICCA\textsubscript{42} classes are devoted to stratocumulus, whose behavior is a key uncertainty in climate projections~\cite{schneider2019possible}.

Preliminary analysis of the AICCA\textsubscript{42} atlas suggests its power for science. Its cloud classes meaningfully group physical properties such as altitude or optical thickness, and also capture distinct textures and patterns. Cloud classes show strikingly different geographical distributions, with distributions evolving seasonally. 
Some classes can be matched to known cloud processes: deep convection in the West Pacific warm pool, for example, or marine stratocumulus decks that form off the west coast of continents. In other cases, cloud classes capture distinctions not previously appreciated, and can lead to new lines of scientific inquiry. We conclude that (1) our methodology has explanatory power, in that it captures regionally unique cloud classes, and (2) 42 clusters is a useful number for a global analysis.


The AICCA approach also opens up possibilities in other areas. 
For example, increasing computing power means the spatial scale of climate simulations has shrunk to the point where their output can resolve complex cloud textures~\cite{norman2022unprecedented}. Unsupervised cloud classification can help in assessing whether models capture those textures correctly. 
More broadly, advances in remote sensing instrumentation mean that many fields have seen large increases in data volume. We have shown here that AI-based methods using a convolutional autoencoder can effectively identify novel patterns in spatial data.
Unsupervised learning offers the possibility of unlocking large satellite datasets and making them tractable for~analysis.

\vspace{6pt} 





{The AICCA data are available upon request until May 1, 2023, on which date all data will be made available without restriction at:~\url{https://github.com/RDCEP/clouds\#download-aicca-dataset} (accessed on 11 November 2022). 
The rotation-invariant autoencoder can be found at: \url{https://acdc.alcf.anl.gov/dlhub/?q=climate} (accessed on 16 September 2022).
Our codebases and Jupyter notebooks are available in Github: \url{https://github.com/RDCEP/clouds} (accessed on 2 November 2022)
} 

\section*{Acknowledgement}
{The authors thank
Rebecca Willett and Davis Gilton for contributing to the RICC design; 
Scott Sinno for providing MODIS products from the NASA Center for Climate Simulation;
NASA for access to MODIS data;
Zhuozhao Li, Sandro Fiore, and Donatello Elia for collaboration and feedback on the AICCA workflow; 
and the Argonne Leadership Computing Facility and the University of Chicago’s Research Computing Center for access to computing resources.
}





\begin{appendices}
\section{}\label{sec:appedix}
\subsection{Rand Index and Adjusted Version for Chance}\label{sec:similarity_scores}

We describe the Rand index used in Section~\ref{sec:stability}.
Let $U = \{U_1, \dots, U_r\}$ and $V = \{V_1,\dots, V_c\}$ be two clustering partitions of a set of
N objects $O = \{o_1, \dots, o_{N_P}\}$,
such that $\bigcup_{i=1}^r U_i = \bigcup_{j=1}^c V_j = O$, and $U_i \cup U_{i^\prime} = \emptyset$ as well as $V_j \cup V_{j^\prime} = \emptyset$ for $1 \leq i \leq r$ and  $1 \leq j \leq c$.
We count how many of the $\binom{N}{2}$ possible pairings of elements in $O$ are in the same or different clusters in \textit{U} and \textit{V}:
\begin{itemize}
    \item $P_{11}$: number of element pairs that are in the \textit{same} clusters in both \textit{U} and \textit{V};
    \item $P_{10}$: number of element pairs that are in \textit{different} clusters in \textit{U}, but in the \textit{same} cluster in \textit{V};
    \item $P_{01}$: number of element pairs that are in the \textit{same} cluster in \textit{U}, but in \textit{different} clusters in \textit{V}; and
    \item $P_{00}$: number of element pairs that are in \textit{different} clusters in both \textit{U} and \textit{V}.
\end{itemize}

The Rand index then computes the fraction of correct cluster assignments:
\begin{equation}\label{eq:randindex}
    \text{RandI}(U,V) = \frac{P_{11} + P_{00}}{P_{11} + P_{10} + P_{01} + P_{00}} = \frac{P_{11} + P_{00}}{\binom{N}{2}}
    \end{equation}
It has value 1 if all pairs of labels are grouped correctly and 0 if none are correct.
The metric is independent of the absolute values of the labels: that is, it allows for permutations.

To illustrate how the Rand index works, consider the two clusterings:
$A = \{d_1\}, \{d_2, d_3\}$ and
$B = \{d_1, d_2 \}, \{ d_3 \}$ of the dataset $D=\{d_1, d_2, d_3\}$.
Here, $N=3$, and there are $\binom{3}{2}$ = 3 possible pairings of the three dataset elements: $(d_1, d_2), (d_1, d_3), (d_2, d_3)$.
Thus:
$P_{11}$ = 0, as no pair is in the same cluster in both $A$ and $B$;
$P_{10}$ = 1, as $d_1$ and $d_2$ are in different clusters in $A$ but the same cluster in $B$;
$P_{01}$ = 1, as $d_2$ and $d_3$ are in different clusters in $A$ but the same cluster in $B$; and 
$P_{00}$ = 1, as $d_1$ and $d_3$ are in different clusters in both $A$ and $B$.
Hence, the Rand index by Equation~(\ref{eq:randindex}) of $A$ and $B$ is (0 + 1)/3 = 0.33.

A difficulty with the Rand index is that its value tends to increase with the number of clusters, hindering comparisons across different numbers of clusters.
In order to permit comparisons of Rand index values across different numbers of clusters, the \textbf{adjusted Rand index} (ARI)~\cite{hubert1985comparing} corrects for co-occurrences due to chance:
   
\begin{equation}\label{eq:ari}
    \text{ARI}(U,V) = \frac{\binom{N}{2}\left( P_{11} + P_{00}\right) - \left[\left( P_{11} + P_{10}\right)\left( P_{11} + P_{01}\right)  + \left( P_{01} + P_{00}\right)\left( P_{10} + P_{00}\right) \right]}{ \binom{N}{2}^{2} -  \left[\left( P_{11} + P_{10}\right)\left( P_{11} + P_{01}\right)  + \left( P_{01} + P_{00}\right)\left( P_{10} + P_{00}\right) \right] },
\end{equation}
where the $P_{xy}$ are as defined above. 

\subsection{Clustering Similarity Test}\label{sec:apdx_clustering_similarity}

We present as Algorithm~\ref{alg:stability1} our implementation of the {\textit{clustering similarity test}}. As described in Section~\ref{sec:stabilitytest1}, we use as the input dataset $D$ all ocean-cloud patches from 2003--2021, inclusive. We define a holdout set, $H$, for evaluation (line~\ref{line:1-1}), and use as our ``perturbed versions'' N subsets selected without replacement from $D\setminus H$ (line~\ref{line:SS-1}). Then, for each number of clusters, $k$, in the range $8\le k\le k_\text{max}$, we:
train RICC on each subset (line~\ref{line:ricc1}); 
apply the trained RICC to generate a clustering for the holdout set (line~\ref{line:ricc2-1}); 
use the adjusted Rand index, ARI, to evaluate pairwise distances between those clusterings (line~\ref{line:mean-1}); 
and average among the 30 clusterings generated by the RICC models $\{ \text{RICC}_k^{i}$, $i \in 1..30 \}$ to determine the mean clustering similarity for that specific cluster number $k$. 
Finally, we calculate the ARI for all $\binom{30}{2} = 435$ combinations of those 30 clusterings and determine the mean ARI score $G_8 .. G_{k_{max}}$ (line~\ref{line:finalscore}).

\begin{algorithm}[h]
\caption{Pseudocode for the clustering similarity test described in Section~\ref{sec:stabilitytest1}.}\label{alg:stability1}
\algnewcommand\algorithmicinput{\textbf{Input:}}
\algnewcommand\algorithmicoutput{\textbf{Output:}}
\algnewcommand\Input{\item[\algorithmicinput]}%
\algnewcommand\Output{\item[\algorithmicoutput]}%

\begin{algorithmic}[1]
\Input $D$: \{ \ocpatches{} for 2003--2021, inclusive \}
\Output $G_8, \dots, G_{k_{\text{max}}}$: Clustering similarity scores for cluster counts from 8 to $k_{max}$.
\vspace{1ex}
\State $H := \{ x \mid x \in D \}$ where $ |H| = N_H$ \Comment{Select holdout set to be used for evaluation} \label{line:1-1}
\For{ i from 1 to N}
    \State Select a subset $S_i := \big\{ x \mid x \in D\setminus H \setminus \bigcup_{j=1}^{\,i-1}S_j \big\}$ with $ |S_i| = N_R$ \label{line:SS-1}
    \For{ k from 8 to $k_{\text{max}}$}
        \State $\texttt{RICC}^{\,i}_{k}$ $\gets$ Train RICC with $k$ clusters on $S_i \cup H $ \label{line:ricc1-1}
        \State $C^{\,i}_{k} \gets \texttt{RICC}^{\,i}_{k} (H)$ \Comment Determine cluster assignments in $H$ with $\texttt{RICC}^{\,i}_{k}$\label{line:ricc2-1}
    \EndFor
\EndFor
\For{ k from 8 to $k_{\text{max}}$ }
    \State $G_{k } = \dfrac{1}{\binom{N}{2}}$ $\mathlarger\sum_{(i,j)\in \binom{N}{2}} \texttt{ARI} \left(C^{\,i}_{k} ,\  C^{\,j}_{k}\right)$ \Comment{Mean similarities for RICC clusters} \label{line:mean-1}
\EndFor
\State Return clustering similarity scores $\{G_8, \dots, G_{k_{max}} \}$  \label{line:finalscore}
\end{algorithmic}
\end{algorithm}


\subsection{Stability Significance Test}\label{sec:apdx_stability_significance}

{Algorithm~\ref{alg:stability} implements the {\textit stability significance test} described in Section~\ref{sec:stabilitytest2}.
For each $k$ in the range 8..$k_\text{max}$, 
we first compute clusterings (line~\ref{line:ricc2}) as in the clustering similarity test of Appendix~\ref{sec:apdx_clustering_similarity} and then compute the mean Rand index score (see Appendix~\ref{sec:similarity_scores}) $G_8 .. G_{k_{max}}$ (line~\ref{line:mean}).  
To produce random label assignments, we first prepare 30 datasets that are sampled from random uniform distributions $\mathcal{U} \in \left[-2\sigma, 2\sigma\right]$ (line~\ref{line:U}). 
We then apply HAC to the random data to generate random labels (line~\ref{line:HAC2}), from which we also calculate the Rand Index for {435} combinations, giving the mean scores $R_8 .. R_{k_{max}}$ (line~\ref{line:mean2}).}
Finally, we compare how the ratio
$\frac{G_k}{R_k}$
varies with number of clusters, $k$ (line~\ref{line:compare2}).

\begin{algorithm}[!h]
\small 
\caption{Pseudocode for the stability significance test described in Section~\ref{sec:stabilitytest2}.}\label{alg:stability}
\algnewcommand\algorithmicinput{\textbf{Input:}}
\algnewcommand\algorithmicoutput{\textbf{Output:}}
\algnewcommand\Input{\item[\algorithmicinput]}%
\algnewcommand\Output{\item[\algorithmicoutput]}%

\begin{algorithmic}[1]
\Input $D$: \{ \ocpatches{} for 2003--2021, inclusive \}, trained rotation-invariant autoencoder $AE$
\Output $\{\frac{G_8}{R_8}, \dots, \frac{G_{k_\text{max}}}{R_{k_\text{max}}} \}$: cluster similarity significance scores
\vspace{1ex}
\State $H := \{ x \mid x \in D \}$ where $ |H| = N_H$ \Comment{Select holdout set to be used for evaluation} \label{line:1}
\State $z = \{ AE(x) : x\in H \}$ \Comment{Use trained autoencoder to compute latent representations}
\State $\sigma = \sqrt{\frac{1}{N_H}}\ \mathlarger\sum_{j=1}^{N_H} \left(z_j - \overline{z}\right)^2 $ \Comment{Calculate standard deviation $\sigma$ for latent representations}
\For{ i from 1 to N}
    \State Select a subset $S_i := \big\{ x \mid x \in D\setminus H \setminus \bigcup_{j=1}^{\,i-1}S_j \big\}$ with $ |S_i| = N_R$ \label{line:SS}
    \State Sample $ U_i := \big\{ u \mid u \in \mathcal{U}\left[-2\sigma, 2\sigma\right]\big\}$ with $ |U_i| =  N_H$, $\mathcal{U}$ a random uniform distribution. \label{line:U}
    \For{ k from 8 to $k_{\text{max}}$}
        \State $\texttt{RICC}^{\,i}_{k}$ $\gets$ Train RICC on $S_i \cup H $ \label{line:ricc1}
        \State $\texttt{RICC}^{\,i}_{k} (H)$ $\gets$ Determine cluster assignments in $H$ \label{line:ricc2}
        \State $\texttt{HAC}^{\,i}_{k}$ $\gets$ Train HAC on $U_i$ \label{line:HAC1}
        \State $\texttt{HAC}^{\,i}_{k} (U_i)$ $\gets$ Determine cluster assignments in $U_i$ \label{line:HAC2}
    \EndFor
\EndFor 

\For{ k from 8 to $k_{\text{max}}$ } \Comment{Calculate averages of cluster similarities}
    \State $G_{k } = \dfrac{1}{\binom{N}{2}}$ $\mathlarger\sum_{(i,j)\in \binom{N}{2}} \left[ {\texttt{RandI}} \left(\texttt{RICC}^{\,i}_{k} (H) ,\  \texttt{RICC}^{\,j}_{k} (H) \right)\right]$ \Comment{Mean similarities for RICC clusters} \label{line:mean}
    \State $R_{k }  = \dfrac{1}{\binom{N}{2}}$ $\mathlarger\sum_{(i,j)\in \binom{N}{2}} \left[{\texttt{RandI}}\left(\texttt{HAC}^{\,i}_{k} (U_i) ,\ 
    \texttt{HAC}^{\,j}_{k}(U_j) \right)\right]$ \Comment{Mean similarities for random clusters} \label{line:mean2}
    \State Calculate $\frac{G_{k}}{R_{k}}$, ratio of stability between RICC and random samples \label{line:ratio} 
\EndFor 
\State Return cluster similarities significance scores, $\{\frac{G_8}{R_8}, \dots, \frac{G_{k_\text{max}}}{R_{k_\text{max}}} \}$ \label{line:compare2}
\end{algorithmic}
\end{algorithm}
\clearpage


\subsection{Mean Patch Occurrence and Trends}\label{sec:jimfigure}
\begin{figure}[hp]
    \centering
    \includegraphics[clip, width=1\textwidth]{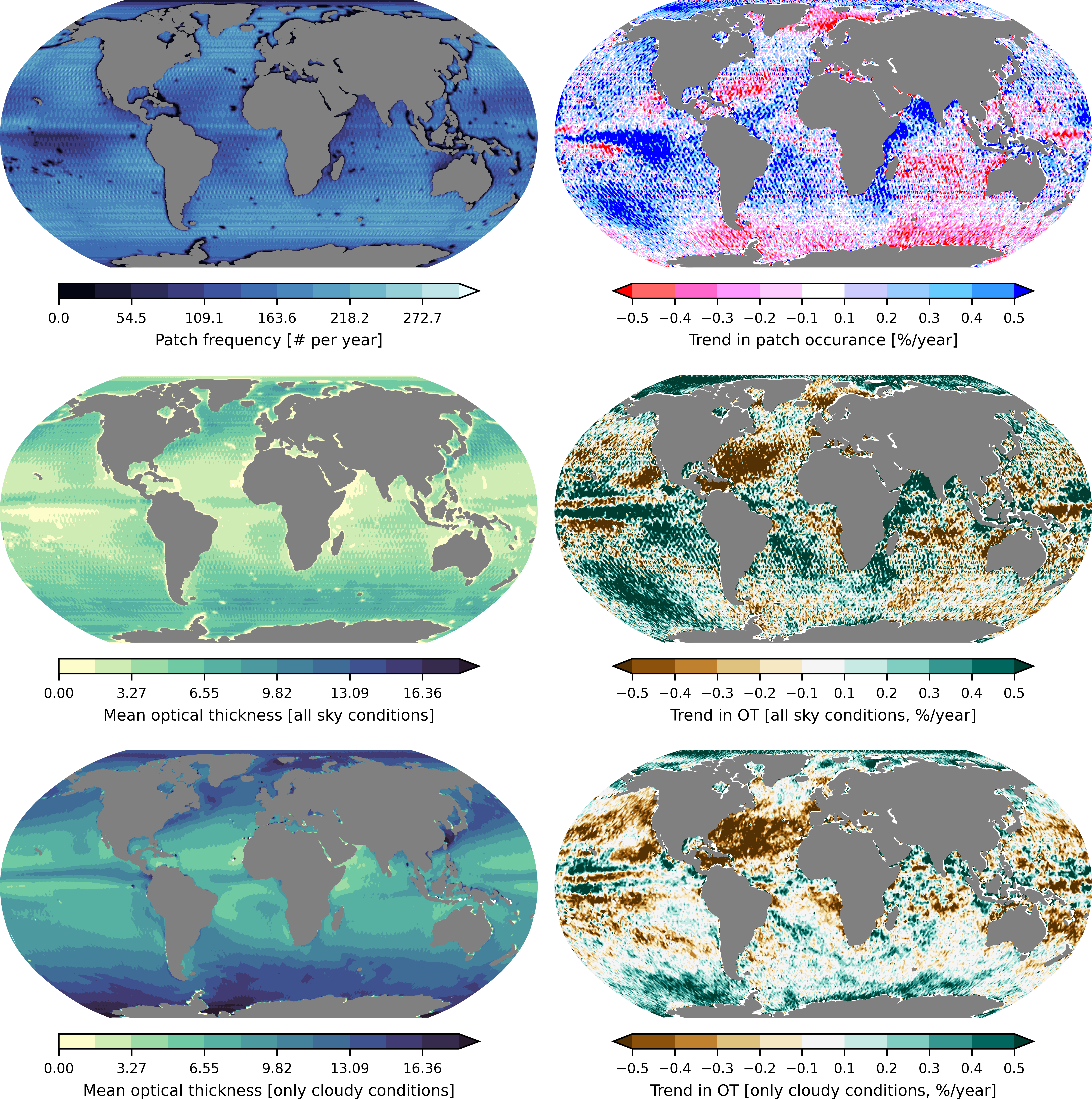}
    \caption{Left column: mean patch occurrence (across all 42 class labels), mean optical thickness in all sky conditions, and mean optical thickness only when identified cloud patches are present. Right column: linear trends in these values shown as a percent change per year compared to the mean value in each grid cell across all years. Data from 2003-2021.}
    \label{fig:jim}
\end{figure}

For summary, we show mean patch occurrence across all 42 classes as well as optical thickness from the mean MODIS proprieties for reference in Figure~\ref{fig:jim}. Linear trends are calculated at the yearly aggregate labels from 2003–2020 in each 1 degree spatial bin.
While not all locations have significant strong trends, we observe that cloudy places are getting more cloudy. Mean
COT and results from trend analysis in Figure~\ref{fig:jim} explain that clouds with larger COT have a positive trend and low
COT have negative trend. Especially, COT and their trends at bottom row in Figure~\ref{fig:jim} tell that under cloudy condition thicker clouds get thicker between 2003–2021.

\end{appendices}

\clearpage
\begin{adjustwidth}{-0cm}{0cm}

\addto\captionsenglish{\renewcommand{\bibname}{References}}



\begin{thebibliography}{999}

\bibitem[Rossow and Schiffer(1991)]{isccp1991}
Rossow, W.B.; Schiffer, R.A.
ISCCP cloud data products.
\emph{Bull.\ Am.\ Meteorol.\ Soc.} 
\textbf{1991}, 
\emph{71},~2--20. 
\url{https://doi.org/10.1175/1520-0477(1991)072<0002:ICDP>2.0.CO;2}.

\bibitem[Rossow {et~al.}(1993)Rossow, Walker, and
  Garder]{rossow1993comparison}
Rossow, W.B.; Walker, A.W.; Garder, L.C.
Comparison of ISCCP and other cloud amounts.
\emph{J.\ Clim.} 
\textbf{1993},
\emph{6},~2394--2418. 
\url{https://doi.org/10.1175/1520-0442(1993)006<2394:COIAOC>2.0.CO;2}.

\bibitem[Rossow and Schiffer(1999)]{rossow1999advances}
Rossow, W.B.; Schiffer, R.A.
Advances in understanding clouds from ISCCP.
\emph{Bull.\ Am.\ Meteorol.\ Soc.}
\textbf{1999},
\emph{80},~2261--2288. \url{https://doi.org/10.1175/1520-0477(1999)080<2261:AIUCFI>2.0.CO;2}.

\bibitem[{World Meteorological Organization}()]{wmo2017}
World Meteorological Organization.
International Cloud Atlas.
Available online: \url{https://cloudatlas.wmo.int/} (accessed on 1 November 2022).

\bibitem[Wood(2012)]{woodStCreview}
Wood, R.
Stratocumulus clouds.
\emph{Mon.\ Weather Rev.} 
\textbf{2012}, 
\emph{140},~2373--2423. \url{https://doi.org/10.1175/MWR-D-11-00121.1}.

\bibitem[Zhang {et~al.}(2018)Zhang, Liu, Zhang, and
  Song]{Zhang2018CloudNetGC}
Zhang, J.; Liu, P.; Zhang, F.; Song, Q.
CloudNet: Ground‐based cloud classification with deep
  convolutional neural network.
\emph{Geophys.\ Res.\ Lett.} 
\textbf{2018},
\emph{45},~8665--8672. \url{https://doi.org/10.1029/2018GL077787}.

\bibitem[Rasp {et~al.}(2019)Rasp, Schulz, Bony, and
  Stevens]{Rasp2019CombiningCA}
Rasp, S.; Schulz, H.; Bony, S.; Stevens, B.
Combining crowd-sourcing and deep learning to understand meso-scale organization of shallow convection.
\emph{ArXiv 1906.01906}
\textbf{2019}.  
\url{https://doi.org/10.48550/arXiv.1906.01906}.

\bibitem[Zantedeschi {et~al.}(2019)Zantedeschi, Falasca, Douglas, Strange,
  Kusner, and Watson-Parris]{Zantedeschi2019CumuloAD}
Zantedeschi, V.; Falasca, F.; Douglas, A.; Strange, R.; Kusner, M.; Watson-Parris, D.
Cumulo: A dataset for learning cloud classes.
\emph{NeurIPS Workshop on Tackling Climate Change with Machine Learning}. 
\textbf{2019}.
Available online: \url{https://www.climatechange.ai/papers/neurips2019/11}  (accessed on 1 November 2022).

\bibitem[Yuan {et~al.}(2020)Yuan, Song, Wood, Mohrmann, Meyer, Oreopoulos,
  and Platnick]{yuan2020applying}
Yuan, T.; Song, H.; Wood, R.; Mohrmann, J.; Meyer, K.; Oreopoulos, L.; Platnick, S.
Applying deep learning to NASA MODIS data to create a community record of marine low-cloud mesoscale morphology.
\emph{Atmos.\ Meas.\ Tech.} 
\textbf{2020},
\emph{13},~6989--6997. \url{https://doi.org/10.5194/amt-13-6989-2020}.

\bibitem[Marais {et~al.}(2020)Marais, Holz, Reid, and
  Willett]{marais2020leveraging}
Marais, W.J.; Holz, R.E.; Reid, J.S.; Willett, R.M. Leveraging spatial textures, through machine learning, to identify aerosols and distinct cloud types from multispectral observations.
\emph{Atmos.\ Meas.\ Tech.}
\textbf{2020},
\emph{13},~5459--5480. \url{https://doi.org/10.5194/amt-13-5459-2020}.

\bibitem[Stevens {et~al.}(2020)Stevens, Bony, Brogniez, Hentgen, Hohenegger, Kiemle, L'Ecuyer, Naumann, Schulz, Siebesma, Vial, Winker, and Zuidema]{stevens2020sugar}
Stevens, B.; Bony, S.; Brogniez, H.; Hentgen, L.; Hohenegger, C.; Kiemle, C.; L'Ecuyer, T.S.; Naumann, A.K.; Schulz, H.; Siebesma, P.A.;  et~al.
Sugar, gravel, fish and flowers: Mesoscale cloud patterns in the trade winds.
\emph{Q.\ J.\ R.\ Meteorol.\ Soc.} 
\textbf{2020}, 
\emph{146},~141--152. 
\url{https://doi.org/10.1002/qj.3662}.

\bibitem[Visa {et~al.}(1998)Visa, Iivarinen, Valkealahti, and Simula]{Visa1995NeuralNB}
Visa, A.; Iivarinen, J.; Valkealahti, K.; Simula, O.
Neural network based cloud classifier. 
In \emph{Industrial Applications of Neural Networks}; World Scientific: Singapore,  
\textbf{1998}; pp.\ 303--309. \url{https://doi.org/10.1142/9789812816955\_0035}.

\bibitem[Tian {et~al.}(1999)Tian, Shaikh, Azimi-Sadjadi, Haar, and Reinke]{Tian1999ASO}
Tian, B.; Shaikh, M.K.; Azimi-Sadjadi, M.R.; Haar, T.H.; Reinke, D.
A study of cloud classification with neural networks using spectral and textural features.
\emph{IEEE Trans.\ Neural Netw.} 
\textbf{1999}, \emph{10}~138--151. \url{https://doi.org/10.1109/72.737500}.

\bibitem[Denby(2020)]{denby2019unsuper}
Denby, L.
Discovering the importance of mesoscale cloud organization through unsupervised classification.
\emph{Geophys.\ Res.\ Lett.}
\textbf{2020},
\emph{47}~e2019GL085190. \url{https://doi.org/10.1029/2019GL085190}.


\bibitem[Kurihana {et~al.}(2019)Kurihana, Foster, Willett, Jenkins, Koenig, Werman, Barros~Lourenco, Neo, and Moyer]{kurihana2019cloud}
Kurihana, T.; Foster, I.T.; Willett, R.; Jenkins, S.; Koenig, K.; Werman, R.; Barros~Lourenco, R.; Neo, C.; Moyer, E.J.
Cloud classification with unsupervised deep learning.
In \emph{Proceedings of the  9th International Workshop on Climate Informatics},
Paris, France, 2--4 October  
\textbf{2019}. 
\url{https://doi.org/10.48550/arXiv.2209.15585}. 


\bibitem[Johnson(1967)]{Johnson1967HAC}
Johnson, S.C.
Hierarchical clustering schemes.
\emph{Psychometrika} 
\textbf{1967},
\emph{32},~241--254. \url{https://doi.org/10.1007/BF02289588}.


\bibitem[Hinton {et~al.}(2011)Hinton, Krizhevsky, and
  Wang]{Hinton2011TransformingA}
Hinton, G.E.; Krizhevsky, A.; Wang, S.
Transforming auto-encoders.
In \emph{International Conference on Artificial Neural Networks}; Springer: Berlin/Heidelberg, Germany, \textbf{2011}; pp.\ 44--51,
\url{https://doi.org/10.1007/978-3-642-21735-7_6}.

\bibitem[Kurihana {et~al.}(2021)Kurihana, Moyer, Willett, Gilton, and Foster]{kurihanaRICC21}
Kurihana, T.; Moyer, E.J.; Willett, R.; Gilton, D.; Foster, I.T.
Data-driven cloud clustering via a rotationally invariant
  autoencoder.
\emph{IEEE Trans.\ Geosci.\ Remote Sens.} 
\textbf{2021},
\emph{60},~4103325. \url{https://doi.org/10.1109/TGRS.2021.3098008}.

\bibitem[Adams(1979)]{Hitchhikers}
Adams, D.
\emph{The Hitchhikers Guide to the Galaxy}; 
Random House: New York, NY, USA, 1979.

\bibitem[{MODIS Characterization Support Team}(2017\natexlab{a})]{myd02}
MODIS Characterization Support Team.
\emph{MODIS/Aqua 1km Calibrated Radiances Product}; Goddard Space Flight Center: Greenbelt, MD, USA, 2017. \url{https://doi.org/10.5067/MODIS/MYD021KM.061}.

\bibitem[{MODIS Characterization Support Team}(2017\natexlab{b})]{mod02}
MODIS Characterization Support Team.
\emph{MODIS/Terra 1km Calibrated Radiances Product}; Goddard Space Flight Center: Greenbelt, MD, USA, 2017. \url{https://doi.org/10.5067/MODIS/MOD021KM.061}.

\bibitem[Rakwatin {et~al.}(2007)Rakwatin, Takeuchi, and
  Yasuoka]{rakwatin2007stripe}
Rakwatin, P.; Takeuchi, W.; Yasuoka, Y.
Stripe noise reduction in MODIS data by combining histogram matching with facet filter.
\emph{IEEE Trans.\ Geosci.\ Remote Sens.}
\textbf{2007},
\emph{45}~1844--1856.
\url{https://doi.org/10.1109/TGRS.2007.895841}.

\bibitem[Rew and Davis(1990)]{rew1990netcdf}
Rew, R.; Davis, G.
NetCDF: An interface for scientific data access.
\emph{IEEE Comput.\ Graph.\ Appl.}
\textbf{1990},
\emph{10},~76--82. 
\url{https://doi.org/10.1109/38.56302}.

\bibitem[Kluyver {et~al.}(2016)Kluyver, Ragan-Kelley, P{\'e}rez, Granger, Bussonnier, Frederic, Kelley, Hamrick, Grout, Corlay, Ivanov, Avila, Abdalla, and Willing]{kluyver2016jupyter}
Kluyver, T.; Ragan-Kelley, B.; P{\'e}rez, F.; Granger, B.; Bussonnier, M.; Frederic, J.; Kelley, K.; Hamrick, J.; Grout, J.; Corlay, S.;  et~al.
Jupyter Notebooks---A publishing format for reproducible
  computational workflows. 
In \emph{Positioning and Power in Academic Publishing: Players, Agents and Agendas}; Loizides, F., Schmidt, B., Eds.; IOS Press: Amsterdam, The Netherlands, 2016: pp.\ 87--90.

\bibitem[Kurihana(2022)]{RICCcode}
Kurihana, T.
Rotation-Invariant Cloud Clustering Code. 2022.
Available online: \url{https://github.com/RDCEP/clouds} (accessed on 1 November 2022).

\bibitem[Chard {et~al.}(2019)Chard, Li, Chard, Ward, Babuji, Woodard, Tuecke, Blaiszik, Franklin, and Foster]{chardDLhub19}
Chard, R.; Li, Z.; Chard, K.; Ward, L.; Babuji, Y.; Woodard, A.; Tuecke, S.; Blaiszik, B.; Franklin, M.J.; Foster, I.T.
DLHub: Model and data serving for science.
In \emph{Proccedings of the IEEE International Parallel and Distributed Processing Symposium,} 
Rio de Janeiro, Brazil, 
20--24 May 2019; pp.\ 283--292. \url{https://doi.org/10.1109/IPDPS.2019.00038}.

\bibitem[Chard {et~al.}(2014)Chard, Tuecke, and Foster]{chard2014efficient}
Chard, K.; Tuecke, S.; Foster, I.T.
Efficient and secure transfer, synchronization, and sharing of big data.
\emph{IEEE Cloud Comput.} 
\textbf{2014}, 
\emph{1},~46--55. 
\url{https://doi.org/10.1109/MCC.2014.52}.

\bibitem[Chard {et~al.}(2020)Chard, Babuji, Li, Skluzacek, Woodard, Blaiszik, Foster, and Chard]{chard2020funcx}
Chard, R.; Babuji, Y.; Li, Z.; Skluzacek, T.; Woodard, A.; Blaiszik, B.; Foster, I.T.; Chard, K.
FuncX: A federated function serving fabric for science.
In \emph{Proceedings of the 29th International Symposium on High-performance Parallel and Distributed Computing}, Stockholm, Sweden, 23--26 June 2020; pp.\ 65--76. \url{https://doi.org/10.1145/3369583.3392683}.

\bibitem[Hinton and Richard(1994)]{Hinton94AEscience}
Hinton, G.E.; Richard, S.Z.
Autoencoders, minimum description length and Helmholtz free energy. 
In \emph{Advances in Neural Information Processing Systems 6};
Morgan-Kaufmann: Burlington, MA, USA,  1994; pp.\ 3--10.

\bibitem[Lee and Seung(1999)]{lee1999learning}
Lee, D.D.; Seung, H.S.
Learning the parts of objects by non-negative matrix factorization.
\emph{Nature} 
\textbf{1999},
\emph{401},~788--791. \url{https://doi.org/10.1038/44565}.

\bibitem[Matsuo {et~al.}(2017)Matsuo, Fukuhara, and
  Shimada]{Matsuo2017TransformIA}
Matsuo, T.; Fukuhara, H.; Shimada, N.
Transform invariant auto-encoder.
In \emph{Proceedings of the IEEE/RSJ International Conference on Intelligent Robots and Systems},
Vancouver, BC, Canada,  24--28 September 2017, 
pp.\ 2359--2364. 
\url{https://doi.org/10.48550/arXiv.1709.03754}.

\bibitem[Nair and Hinton(2010)]{Nair2010RectifiedLU}
Nair, V.; Hinton, G.E.
Rectified linear units improve restricted Boltzmann machines.
In \emph{Int.\ Conf.\ Mach.\ Learn.}, 2010.
Available online: \url{https://icml.cc/Conferences/2010/papers/432.pdf} (accessed on 1 November 2022).

\bibitem[Ioffe and Szegedy(2015)]{Ioffe2015BatchNA}
Ioffe, S.; Szegedy, C.
Batch normalization: Accelerating deep network training by reducing internal covariate shift.
\emph{Int.\ Conf.\ Mach.\ Learn.}
\textbf{2015}, 
\emph{37},~448--456. \url{https://doi.org/10.48550/arXiv.1502.03167}.

\bibitem[Ward~Jr(1963)]{Ward1963HierarchicalGT}
Ward~Jr, J.H.
Hierarchical grouping to optimize an objective function.
\emph{J.\ Am.\ Stat.\ Assoc.} {\bf 1963} \emph{58},~236--244. \url{https://doi.org/10.1080/01621459.1963.10500845}.

\bibitem[Jenkins {et~al.}(2019)Jenkins, Moyer, Foster, Kurihana, Willett, Maire, Koenig, and Werman]{Jenkins2019DevelopingUL}
Jenkins, S.; Moyer, E.J.; Foster, I.T.; Kurihana, T.; Willett, R.; Maire, M.; Koenig, K.; Werman, R.
Developing unsupervised learning models for cloud classification.
\emph{AGU Fall Meet. Abstr.}
\textbf{2019}, 
A51U-2673.

\bibitem[Moertini {et~al.}(2018)Moertini, Suarjana, Venica, and Karya]{moertini2018big}
Moertini, V.S.; Suarjana, G.W.; Venica, L.; Karya, G.
Big data reduction technique using parallel hierarchical agglomerative clustering.
\emph{IAENG Int. J. Comput. Sci.}
\textbf{2018},
\emph{45},~1

\bibitem[Varoquaux {et~al.}(2015)Varoquaux, Buitinck, Louppe, Grisel, Pedregosa, and Mueller]{varoquaux2015scikit}
Varoquaux, G.; Buitinck, L.; Louppe, G.; Grisel, O.; Pedregosa, F.; Mueller, A.
Scikit-learn: Machine learning without learning the machinery.
\emph{Getmobile Mob. Comput. Commun.} 
\textbf{2015},
\emph{19},~29--33. \url{https://doi.org/10.1145/2786984.2786995}.


\bibitem[Jin {et~al.}(2015)Jin, Liu, Chen, Hendrix, Agrawal, and Choudhary]{jin2015scalable}
Jin, C.; Liu, R.; Chen, Z.; Hendrix, W.; Agrawal, A.; Choudhary, A.
A scalable hierarchical clustering algorithm using Spark.
In \emph{Proceedings of the IEEE First International Conference on Big Data Computing Service and Applications}, Redwood City, CA, USA, 30 March 2015--2 April 2015; pp.\ 418--426. \url{https://doi.org/10.1109/BigDataService.2015.67}.

\bibitem[Sumengen {et~al.}(2021)Sumengen, Rajagopalan, Citovsky, Simcha, Bachem, Mitra, Blasiak, Liang, and Kumar]{sumengen2021scaling}
Sumengen, B.; Rajagopalan, A.; Citovsky, G.; Simcha, D.; Bachem, O.; Mitra, P.; Blasiak, S.; Liang, M.; Kumar, S.
Scaling hierarchical agglomerative clustering to billion-sized datasets. \emph{ArXiv} \textbf{2021}, 
\url{https://doi.org/10.48550/arXiv.2105.11653}.

\bibitem[Monath {et~al.}(2021)Monath, Dubey, Guruganesh, Zaheer, Ahmed, McCallum, Mergen, Najork, Terzihan, Tjanaka, Wang, and Wu]{monath2021scalable}
Monath, N.; Dubey, K.A.; Guruganesh, G.; Zaheer, M.; Ahmed, A.; McCallum, A.; Mergen, G.; Najork, M.; Terzihan, M.; Tjanaka, B.;  et~al.
Scalable hierarchical agglomerative clustering.
In \emph{Proceedings of the 27th ACM SIGKDD Conference on Knowledge Discovery and Data Mining}, Singapore, 14--18 August 2021; pp.\ 1245--1255. \url{https://doi.org/10.1145/3447548.3467404}.

\bibitem[Babuji {et~al.}(2019)Babuji, Woodard, Li, Katz, Clifford, Kumar, Lacinski, Chard, Wozniak, Foster, et~al.]{babuji2019parsl}
Babuji, Y.; Woodard, A.; Li, Z.; Katz, D.S.; Clifford, B.; Kumar, R.; Lacinski, L.; Chard, R.; Wozniak, J.M.; Foster, I.T.;  et~al.
Parsl: Pervasive parallel programming in Python.
In \emph{Proceedings of the 28th International Symposium on High-Performance Parallel and Distributed Computing},
Phoenix, AZ, USA, 22--29 June 2019; pp.\ 25--36. \url{https://doi.org/10.1145/3307681.3325400}.

\bibitem[Santos and Embrechts(2009)]{santos2009use}
Santos, J.M.; Embrechts, M.
On the use of the adjusted Rand index as a metric for evaluating supervised classification. 
In \emph{International Conference on Artificial Neural Networks}; Springer: Berlin/Heidelberg, Germany, 2009; pp.\ 175--184. 
\url{https://doi.org/10.1007/978-3-642-04277-5\_18}.

\bibitem[Von~Luxburg(2010)]{von2010clustering}
Von~Luxburg, U.
\emph{Clustering Stability: An Overview}; Now Publishers Inc.:  Norwell, MA, USA, 2010.
\url{https://doi.org/10.48550/arXiv.1007.1075}.

\bibitem[Tselioudis {et~al.}(2013)Tselioudis, Rossow, Zhang, and Konsta]{tselioudis2013global}
Tselioudis, G.; Rossow, W.; Zhang, Y.; Konsta, D.
Global weather states and their properties from passive and active satellite cloud retrievals.
\emph{J. Clim.}
\textbf{2013},
\emph{26},~7734--7746. \url{https://doi.org/10.1175/JCLI-D-13-00024.1}.


\bibitem[McDonald and Parsons(2018)]{mcdonald2018comparison}
McDonald, A.; Parsons, S.
A comparison of cloud classification methodologies: Differences between cloud and dynamical regimes.
\emph{J. Geophys. Res. Atmos.} 
\textbf{2018},
\emph{123},~11--173. \url{https://doi.org/10.1029/2018JD028595}.


\bibitem[Schuddeboom {et~al.}(2018)Schuddeboom, McDonald, Morgenstern, Harvey, and Parsons]{schuddeboom2018regional}
Schuddeboom, A.; McDonald, A.J.; Morgenstern, O.; Harvey, M.; Parsons, S.
Regional regime-based evaluation of present-day general circulation model cloud simulations using self-organizing maps.
\emph{J. Geophys. Res. Atmos.} 
\textbf{2018},
\emph{123},~4259--4272. \url{https://doi.org/10.1002/2017JD028196}.


\bibitem[Bholowalia and Kumar(2014)]{bholowalia2014ebk}
Bholowalia, P.; Kumar, A.
EBK-means: A clustering technique based on elbow method and k-means in WSN.
\emph{Int. J. Comput. Appl.} 
\textbf{2014},
\emph{105},~17--24.

\bibitem[Rousseeuw(1987)]{rousseeuw1987silhouettes}
Rousseeuw, P.J.
Silhouettes: A graphical aid to the interpretation and validation of cluster analysis.
\emph{J. Comput. Appl. Math.}
\textbf{1987},
\emph{20},~53--65. \url{https://doi.org/10.1016/0377-0427(87)90125-7}.

\bibitem[Tibshirani {et~al.}(2001)Tibshirani, Walther, and
  Hastie]{tibshirani2001estimating}
Tibshirani, R.; Walther, G.; Hastie, T.
Estimating the number of clusters in a data set via the gap statistic.
\emph{J. R. Stat. Soc. Ser. Stat. Methodol.}
\textbf{2001},
\emph{63},~411--423. \url{https://doi.org/10.1111/1467-9868.00293}.

\bibitem[Jin {et~al.}(2017)Jin, Oreopoulos, and Lee]{jin2017simplified}
Jin, D.; Oreopoulos, L.; Lee, D.
Simplified ISCCP cloud regimes for evaluating cloudiness in CMIP5 models.
\emph{Clim. Dyn.}
\textbf{2017},
\emph{48},~113--130. \url{https://doi.org/10.1007/s00382-016-3107-6}.

\bibitem[clo()]{cloudtypes}
ISCCP Definition of Cloud Types.
Available online: \url{https://isccp.giss.nasa.gov/cloudtypes.html} (accessed on 1 May 2022).

\bibitem[Gumley {et~al.}(2003)Gumley, Descloitres, and
  Schmaltz]{gumley2003creating}
{Gumley, L.; Descloitres, J.; Schmaltz, J.
\emph{Creating Reprojected True Color MODIS Images: A Tutorial};
University of Wisconsin---Madison: Madison, WI, USA, 2003,
Volume 19.

\bibitem[Riggs {et~al.}(2015)Riggs, Hall, and Rom{\'a}n]{riggs2015modis}
Riggs, G.A.; Hall, D.K.; Rom}{\'a}n, M.O.
\emph{MODIS Snow Products Collection 6 User Guide};
National Snow and Ice Data Center: Boulder, CO, USA, 2015, 
Volume 66.
Available online:
\url{https://modis-snow-ice.gsfc.nasa.gov/uploads/C6\_MODIS\_Snow_User\_Guide.pdf} (accessed on 1 November 2022).

\bibitem[Schneider {et~al.}(2019)Schneider, Kaul, and
  Pressel]{schneider2019possible}
Schneider, T.; Kaul, C.M.; Pressel, K.G.
Possible climate transitions from breakup of stratocumulus decks under greenhouse warming.
\emph{Nat. Geosci.} 
\textbf{2019},
\emph{12}~163--167. \url{https://doi.org/10.1038/s41561-019-0310-1}.

\bibitem[Norman {et~al.}(2022)Norman, Bader, Eldred, Hannah, Hillman, Jones,
  Lee, Leung, Lyngaas, Pressel, Sreepathi, Taylor, and
  Yuan]{norman2022unprecedented}
Norman, M.R.; Bader, D.A.; Eldred, C.; Hannah, W.M.; Hillman, B.R.; Jones, C.R.; Lee, J.M.; Leung, L.; Lyngaas, I.; Pressel, K.G.; et~al.
Unprecedented cloud resolution in a GPU-enabled full-physics atmospheric climate simulation on OLCF’s Summit supercomputer.
\emph{Int. J. High Perform. Comput. Appl.} 
\textbf{2022}, 
\emph{36},~93--105. \url{https://doi.org/10.1177/10943420211027539}.

\bibitem[Hubert and Arabie(1985)]{hubert1985comparing}
Hubert, L.; Arabie, P.
Comparing partitions.
\emph{J. Classif.} \textbf{1985},
\emph{2},~193--218. \url{https://doi.org/10.1007/BF01908075}.

\end{thebibliography}

\end{adjustwidth}
\end{document}